\documentclass[twocolumn,showpacs,preprintnumbers,amsmath,amssymb,superscriptaddress]{revtex4}
\usepackage{graphicx}
\usepackage{epsfig}
\usepackage{color}
\usepackage{dcolumn}% Align table columns on decimal point
\usepackage{bm}% bold math
\usepackage{amsmath,amssymb,amsthm}

\newcommand{\KAG}{Graduate School of Science and Engineering, Kagoshima University, 1-21-35 Korimoto, Kagoshima 890-0065, Japan}
\newcommand{\PRE}{PRESTO, JST, 4-1-8 Honcho Kawaguchi-shi, Saitama 332-0012, Japan}
\newcommand{\FO}{RIKEN Advanced Science Institute, 2-1 Hirosawa, Wako-shi, Saitama 351-0198, Japan}
\newcommand{\IIS}{Institute of Industrial Science, The University of Tokyo, 4-6-1 Komaba, Meguro-ku, Tokyo 153-8505, Japan}

\newcommand{\BSI}{RIKEN Brain Science Institute, 2-1 Hirosawa, Wako-shi, Saitama 351-0198, Japan}
\newcommand{\PIK}{Potsdam Institute for Climate Impact Research,  D-14412 Potsdam, P.O. Box 601203, Germany}
\newcommand{\HUM}{Institute of Physics, Humboldt University, D-12489 Berlin, Germany}
\newcommand{\ABE}{Institute for Complex Systems and Mathematical Biology, University of Aberdeen, Aberdeen AB24 3UE, United Kingdom}
\begin{document}
\title{Chaotic Phase Synchronization in Bursting-neuron Models \\Driven by a Weak Periodic Force}

\author{Hiroyasu Ando}\affiliation{\BSI}
\author{Hiromichi Suetani}\affiliation{\KAG}\affiliation{\PRE}\affiliation{\FO}
\author{J\"urgen Kurths}\affiliation{\PIK}\affiliation{\HUM}\affiliation{\ABE}
\author{Kazuyuki Aihara}\affiliation{\IIS}

\date{\today}
\begin{abstract}
We investigate {the} entrainment of  
a neuron model {exhibiting a} chaotic spiking-bursting behavior
in response  
to a weak periodic 
force.   
{This model exhibits two 
types of oscillations with different characteristic time scales, 
namely, long and short time scales.}  
Several {types} of phase synchronization 
are observed, such as $1\ :\ 1$ phase locking between  a single spike 
and one period of the force and $1\ :\ l$ phase locking between 
 the period of slow oscillation underlying bursts and 
$l$ periods of the force. 
Moreover,  spiking-bursting oscillations with 
chaotic firing patterns can be synchronized with the periodic force.   
Such {a type of} phase synchronization is
detected from the {position} of a set of points on a unit circle, which is 
determined by the phase of the periodic force at each {spiking time}. 
We show that this detection method is effective for a system with multiple time scales. 
{Owing} to the existence of both {the} short and {the} 
long time scales, 
 two characteristic phenomena are found around 
the transition point to chaotic phase synchronization.    
One {phenomenon shows} that the average time interval between successive phase slips 
exhibits  a power-law scaling against the {driving force} strength 
and {that} the scaling exponent has an unsmooth dependence on 
{the} changes in the 
{driving force} strength.  
The other {phenomenon shows} that Kuramoto's order parameter  before the transition 
exhibits stepwise behavior as a function of the {driving force} strength,  
contrary to 
the smooth transition in a model with a single time scale.   
\end{abstract}
\maketitle

\section{Introduction}
 Since the discovery of synchronization in pendulum 
clocks by Huygens, synchronous behavior has been widely observed not only  
in physical systems but also in biological ones {such as}
pacemaker cells in the heart, chirps of crickets, and 
fetal-maternal heart rate synchronization \cite{vanLeeuwen09}.
Such synchronization phenomena have been  
studied theoretically in terms of nonlinear dynamics, 
{particularly} by exploiting oscillator models  
\cite{PikovskyBook}. 
For example, synchronization observed in  
fireflies can be modeled {using} nonlinear 
periodic oscillators and {is} described as {\it phase synchronization}. 
Further, it has been {indicated} that the notion of phase 
synchronization can be extended to 
chaotic oscillators. This phenomenon is called 
{\it chaotic phase
  synchronization} (CPS) \cite{RosenblumPRL96,PikovskyPhysD1997,BKOVZ}. 

Furthermore, synchronization phenomena in neural systems have also 
attracted {considerable} attention in recent years.  
At the macroscopic level 
of {the} brain activity, synchronous behavior 
has been observed in electroencephalograms, 
local field potentials, etc. 
These observations raise a possibility that such neural synchronization 
plays an important role in brain functions {such as} perception \cite{Varela99} 
{as well as} even in dysfunctions such as Parkinson's disease and epilepsy \cite{Tass98,Hammond07,VarelaPhysD99, GrassbergerPhysD99}.  
 In addition, at the level of a single neuron, 
it has been observed that 
specific spiking-bursting neurons in the cat visual cortex 
contribute to the synchronous activity evoked by visual 
stimulation \cite{GrayMcCormick96}{; further,}  
 in animal models of Parkinson's disease{,} 
several {types} of {bursting} neurons     
are synchronized  \cite{Hammond07}. 
Moreover, two coupled neurons extracted 
from the central pattern generator of the stomatogastric ganglion in a lobster exhibit synchronization with irregular spiking-bursting behavior  \cite{ElsonPRL98, Szucs01}.

  Hence, it is important to 
use mathematical models of neurons to
  examine the mechanism of 
 neuronal synchronization with spiking-bursting behavior.   
   As mathematical models 
 that include such neural oscillations, the Chay model \cite{Chay} and the 
 Hindmarsh-Rose (HR) model \cite{HR} have been widely used. 
 These models can generate both regular and chaotic bursting on the basis of  
 {\it slow-fast} dynamics.  The slow and fast dynamics 
 correspond to slow oscillations surmounted by spikes   
 and spikes within each burst, respectively. The former is 
 related to a long time scale{,} and the latter{,} to a short one.

  {Phase} synchronization in such neuronal models is different from that in  
  ordinary chaotic systems {such as} the R\"ossler system, {owing to 
  the fact that neuronal models typically exhibit} multiple 
  time scales. However, it is possible to {quantitatively} analyze the neuronal models   
  by simplification{, for example, by} reducing the number of phase variables 
  to {1} by a projection {of an attractor (a projection} onto a delayed coordinate and/or a velocity space   
  \cite{ShuaiPLA99,PereiraPLA07}{)}.
Recently, a method  called {\it localized sets} {technique} has been proposed for detecting 
 phase synchronization in neural networks{,} without explicitly 
defining the phase \cite{PBKpre,BaptistaPRE08,PereiraEPJ07}.

In this paper, 
we {focus on synchronization in} periodically driven {single} bursting neuron models, {which is simpler than that in a network of neurons}. 
In previous studies,  
 phase synchronization of such a neuron  
with a driving force has been considered both theoretically 
\cite{IvanchenkoPRL04} and experimentally \cite{Szucs01}. 
In these studies, the period of the driving force was  
made close to that of the slow oscillation of a driven neuron. 
On the other hand,  in this work,  
we adopt the Chay model \cite{Chay} to investigate whether phase synchronization also occurs 
with {the application of} a force whose period is as short as that of the spikes.  
In particular, we focus on the effect of the slow mode 
(slow oscillation) on the synchronization of the fast mode (spikes).

 {It should be noted} that this fast driven system may be significant from the viewpoint of 
neuroscience. In fact, fast oscillations {with} local field potentials have been 
observed in the hippocampus and are correlated with synchronous activity at the 
level of a single neuron \cite{BuzsakiBook06}.

{ {From} intensive numerical simulations of our model},   
{we find that} the localized sets technique {can be used to} 
detect {CPS} 
between the spikes and the periodic driving force, 
even {in the case of} multiple time scales.
Furthermore, we find two 
characteristic properties around {the} transition point to {CPS}. 
First, the average time interval 
between successive phase slips exhibits 
 a power-law scaling against the {driving force} strength.  
The scaling exponent undergoes an unsmooth change as the {driving force} strength is varied.  
Second, an order parameter $R$, which measures 
the degree of phase synchronization, {shows} a stepwise 
dependence on the {driving force} strength $K$ before the transition. 
That is,  $R(K)$ does not 
 increase monotonically with $K$ but includes a plateau  
over a range of $K$  (a step){,} where $R$ is almost constant. 
Both of these characteristics are {attributed} to the effects of the slow mode 
on the fast mode 
and have not been observed in a system with a single time scale.

This paper is organized as follows. Section 
\ref{model} explains the model and {describes} an analysis method for 
spiking-bursting oscillations. Section \ref{result} {presents} the results 
{of this study}.  
Finally, Section \ref{summary} 
 summarizes our results and discusses their neuroscientific 
significance with a view to future work.

{\section{Bursting Neuron Model}\label{model}}
As an illustrative example of a bursting neuron model,  
we consider the model proposed by Chay, which is a 
Hodgkin-Huxley{-}type conductance-based model {expressed} as follows \cite{Chay}:
\begin{eqnarray}
\nonumber \dot{V}&=&g^*_{\text I}m^3_{\infty}h_{\infty}(V_{\text{I}}-V)+g_{\text{K,V}}^*{q}^4(V_{\text{K}}-V)\\
              & &+g_{\text{K,C}}^* \frac{C}{1+C}(V_{\text{K}}-V)+g_{\text{L}}^*(V_{\text{L}}-V),\label{dVdt}\\
\dot{{q}}&=&({q}_{\infty}-{q})/\tau_{{q}},\label{dndt}\\
\dot{C}&=&\rho[m_{\infty}^3h_{\infty}(V_{\text{C}}-V)-k_{\text{C}}C].\label{dCdt}                     
\end{eqnarray}

Equation (\ref{dVdt}) represents the dynamics of the membrane potential
$V$, where $V_{\text{I}}$, $V_{\text{K}}$, and $V_{\text{L}}$ 
are the reversal potentials for mixed 
Na$^+$ and Ca$^{2+}$ ions, K$^+$ ions, and the leakage current, respectively. 
The concentration 
of {the} intracellular Ca$^{2+}$ ions divided by its dissociation constant 
from the receptor is denoted by $C$.  
The maximal conductances divided 
by the membrane capacitance 
are denoted by 
$g_{\text{I}}^*$, $g_{\text{K,V}}^*$, $g_{\text{K,C}}^*$, and $g_{\text{L}}^*$, 
where subscripts (I), (K, V), (K, C), 
and (L) refer to the voltage-sensitive mixed ion channel, 
voltage-sensitive K$^+$ channel, Ca$^{2+}$-sensitive K$^+$ channel, and leakage current, respectively.
Finally, $m_{\infty}$ and $h_{\infty}$ are the probabilities of activation 
and inactivation of the mixed channel, respectively. 

In Eq. (\ref{dndt}), the dynamical variable ${q}$ {denotes}  
the probability of opening the voltage-sensitive K$^+$-channel, 
where $\tau_{q}$ is the relaxation time (in seconds), and ${q}_{\infty}$ is 
the steady{-}state value of ${q}$. 

{It should be noted} that, {on the basis of} the formulation in \cite{Chay}, 
{the} variables $m_{\infty}$, 
$h_{\infty}$, and ${q}_{\infty}$ are  
described by
$y_{\infty}=\alpha_y/(\alpha_y+\beta_y),$ where $y$ stands for 
$m,\ h$, or ${q}$ with 
\begin{eqnarray*}
\alpha_m&=&0.1(25+V)/[1-\exp(-0.1V-2.5)],\\
\beta_m&=&4\exp[-(V+50)/18],\\
\alpha_h&=&0.07\exp(-0.05V-2.5),\\
\beta_h&=&1/[1+\exp(-0.1V-2)],\\
\alpha_{q}&=&0.01(20+V)/[1-\exp(-0.1V-2)],\\
\beta_{q}&=&0.125\exp[-(V+30)/80].
\end{eqnarray*}
{Further}, $\tau_{q}$ is defined as 
\begin{eqnarray*}
\tau_{q}=[230(\alpha_{q}+\beta_{q})]^{-1}.
\end{eqnarray*}

In Eq. (\ref{dCdt}),  $k_{\text{C}}$, 
$\rho$, and $V_{\text{C}}$ are the efflux rate constant 
of {the} intracellular Ca$^{2+}$ 
ions, a proportionality constant, and the reversal potential for 
Ca$^{2+}$ ions, respectively.

In this study, a sinusoidal driving force  
with 
amplitude $K$ and frequency $f$ is added in Eq. (\ref{dVdt}) as follows:
\begin{align*}
\nonumber \dot{V}&= g^*_{\text{I}}m^3_{\infty}h_{\infty}(V_{\text{I}}-V)+g_{\text{K,V}}^*{q}^4(V_{\text{K}}-V)\\
\nonumber              & +g_{\text{K,C}}^* \frac{C}{1+C}(V_{\text{K}}-V)+g_{\text{L}}^*(V_{\text{L}}-V)\\
                       & +K \sin 2 \pi f t. \tag{1'}\label{dVdt'}
\end{align*}
In the following sections, we {fix} the frequency at {$f=f_0$}  and 
vary {the} amplitude $K$ 
 to investigate the response of the system.

\begin{figure}[htbp]
 \begin{minipage}{.45\textwidth}
  \begin{center}
    \includegraphics[keepaspectratio=true,height=.5\textwidth]{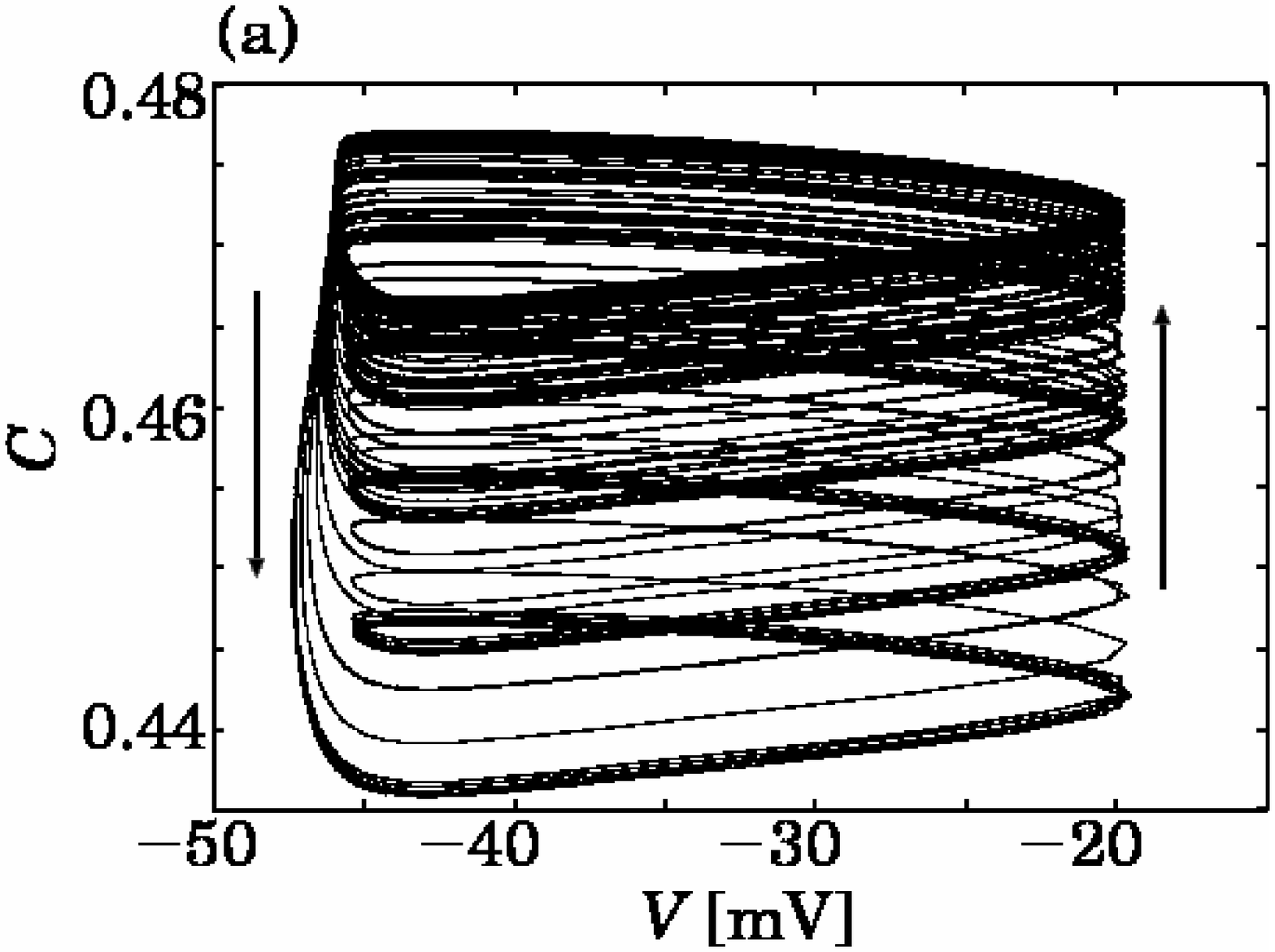}
 \end{center}
    \begin{center}
    \rotatebox{0}{\includegraphics[keepaspectratio=true,height=.5\textwidth]{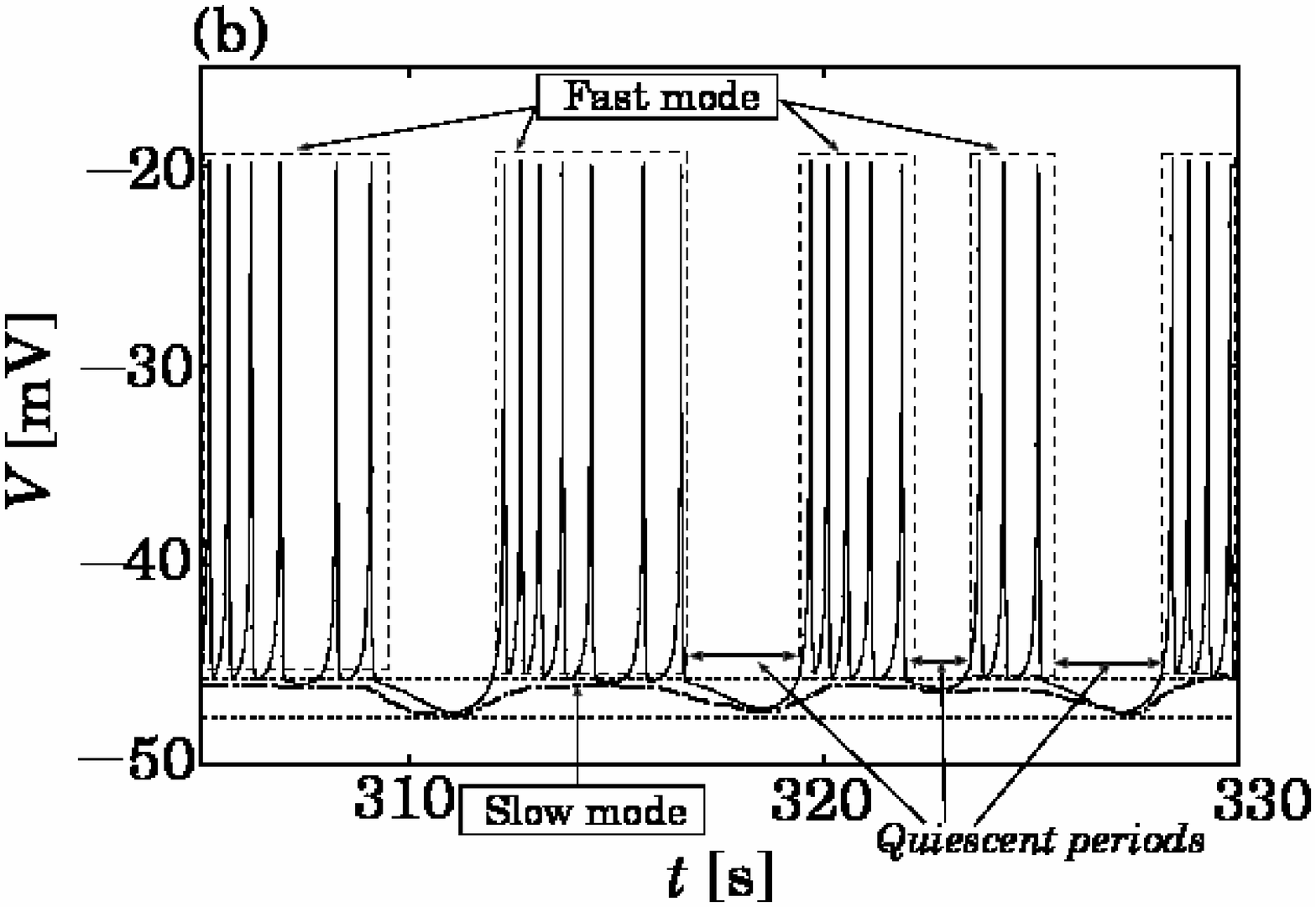}}
    \end{center}
\begin{center}
    \includegraphics[keepaspectratio=true,height=.5\textwidth]{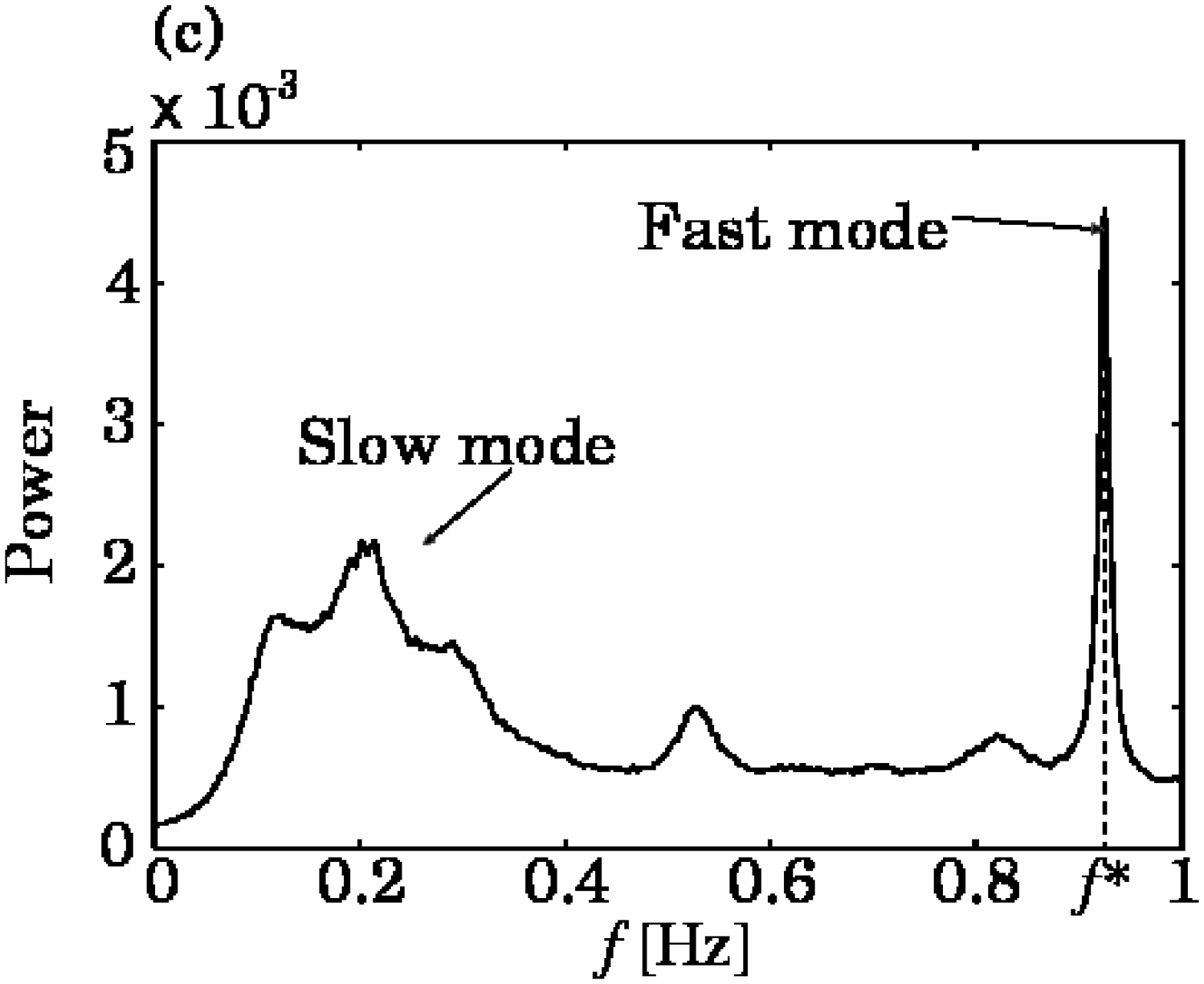}
  \end{center}
   \end{minipage}
  \caption{(a) {Chaotic} attractor in {the} $(V,C)$ space. The arrows 
  indicate the direction of the trajectory. (b) Time series of $V(t)$ 
   for the chaotically bursting behavior described by Eqs. (1)--(3). 
  The terms {``}fast mode{''}  and {``}slow mode{''} describe the spiking 
  oscillation and the oscillation in the lower envelope of $V(t)$ between 
  the dotted lines (shown by the dashed-dotted curve), respectively. 
  (c) Power spectrum for time series of $V(t)$. The spectrum is averaged 
  over 100 time series with a length of $2^{16}\times \Delta t$, 
  where $\Delta t =0.005$ s is the time step for 
  the numerical integration.}
  \label{fig:Fig_attractor.eps}
\end{figure}

The values of the reversal potentials and {the} fixed parameters used
in our simulation are {listed} in Table \ref{parameter}. The value of  
$g_{\text{K,C}}^*$  {significantly} influences the dynamics 
of the system, and {a} chaotically 
bursting behavior can be {observed} in the vicinity of  
$g_{\text{K,C}}^*=11$. We use this value in our simulation. 
Figure \ref{fig:Fig_attractor.eps} shows (a) the chaotic attractor,  
(b) the time series of $V(t)$, and (c) the average power spectrum 
of the time series for $K=0$, where 
the time series {includes} two time scales {---} one for the spikes 
within  each burst and {the other one} for the bursts themselves.

For discussions in the rest of this paper, we introduce the following 
terminology for the time series of the Chay model: the fast mode 
describes the spiking oscillation  
 in the dashed rectangles in Fig. \ref{fig:Fig_attractor.eps}(b), 
 {whereas} the slow mode describes the oscillation  
 in the lower envelope of $V(t)$ between the dotted lines,   
where the dashed-dotted curve shows the slow oscillation.  
{The variable} $C$  dominates  the slow dynamics with {the} time 
constant $1/\rho$. In fact, the {decrease in} $V$ (i.e., hyperpolarization) 
between the bursts {shown} in Fig. \ref{fig:Fig_attractor.eps}(b) corresponds to 
the  {decrease in} $C$ {shown} in Fig. \ref{fig:Fig_attractor.eps}(a).  
{On the other hand,} the {increase in} $C$ {shown} in Fig. \ref{fig:Fig_attractor.eps}(a)  
corresponds to 
the spiking of $V$  in Fig. \ref{fig:Fig_attractor.eps}(b).  
 Hereafter,  
{the period for}  
hyperpolarization of $V$ is called {the} {\it {quiescent period}}, as indicated in 
Fig. \ref{fig:Fig_attractor.eps}(b).   {It should be {noted} that the amplitude 
of $K$ takes small values in our simulation {as} compared with the change {in} voltage 
for a spike, {such} that the 
driving force is weak.}

A clear peak can be observed in the high-frequency part of the power 
spectrum (Fig. \ref{fig:Fig_attractor.eps}(c)), corresponding to the fast spiking 
activity. Additionally, {a} broadband peak for the slow oscillation is  
observed in the low-frequency part. In what follows, let us investigate 
the system under {force} with {a} frequency close to that of the spiking mode  
(i.e., $f=f^* \ {\approx} \ 0.924$ {Hz}), which is the natural frequency of the 
fast dynamics of the system.

\begin{table}[htbp]
 \caption{Parameters used in the numerical simulations.}\label{parameter}
 \begin{center}
  \begin{tabular}{|c |c| c|}
  \hline
    Parameter  &  Value & Unit \\
    \hline
     $V_K$  &  $-75$  & mV   \\
    \hline
     $V_I$  & $100$   & mV   \\
    \hline
     $V_L$  & $-40$   & mV   \\
    \hline
     $V_C$  & $100$   & mV   \\
    \hline
     $V_n$  & $-30$   & mV   \\
    \hline
     $V_m$  & $-50$   & mV   \\
    \hline
     $g_{K,V}^*$  &  $1700$  & s$^{-1}$   \\
    \hline
     $g_I^*$  &  $1800$  &  s$^{-1}$  \\
    \hline
     $g_L^*$  &  $7$   &  s$^{-1}$  \\
    \hline
     $k_C$  &  $3.3/18$  & mV   \\
    \hline
     $\rho$  & $0.27$   & mV$^{-1}$s$^{-1}$   \\
    \hline
  \end{tabular}
 \end{center}
\end{table}

\begin{figure}[htbp]
  \begin{center}
    \rotatebox{-90}{\includegraphics[keepaspectratio=true,height=.5\textwidth]{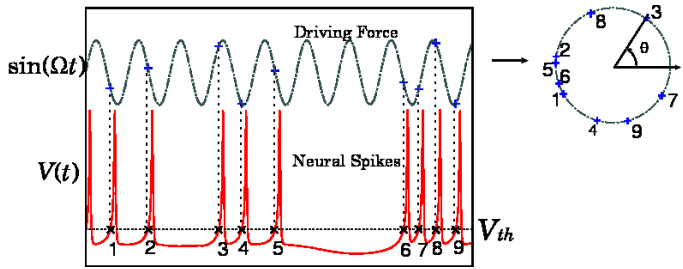}}
% \rotatebox{-90}{\includegraphics[keepaspectratio=true,height=100mm]{schematic_phasesets.eps}}
  \end{center}
  \caption{(Color online) Schematic illustration of {\it {spiking time} points} 
  (STPs). {Under} the condition of upward crossing, 
  the relevant threshold $V_{th}$ for the spikes ({bottom left})  
  is indicated by the  horizontal dotted line. 
  In crossing the threshold{,} the numbered crosses ($\times$) 
   correspond to the {plus symbols} (+) 
  on the sinusoidal driving force ({top left}). The {plus symbols} 
  on the circle (right) 
  correspond to those {on the top left}. The {plus symbols}  
  on the circle are the phases of the {STPs}.}
  \label{fig:schematic_phasesets.eps}
\end{figure}

Next, in order to investigate phase synchronization between the spikes 
and the periodic driving force, we consider {the} phase 
$\theta_n$ of the driving force at $t=t_n$ when 
the $n$th spike occurs in the neuron 
(see \cite{PikovskyChaos97}). 
Figure \ref{fig:schematic_phasesets.eps} illustrates {the manner in which}  
 the phase variable of the driving force  at $t_n$, 
which is defined as the moment when 
$V(t)$ exceeds a certain threshold $V_{th}${, can be measured}.
Once a sequence of  $t_n$ is determined, we can assign points  
$\theta_n$ on a unit circle, where each $\theta_n$ is determined as the phase of the sinusoidal 
force at time $t_n$. {We assume that $0\le \theta_n < 2\pi$.} 
We {term these points the} {\it {spiking time} points} (STPs). 

We can detect {the} CPS between the driven system and the driving force 
as a localization of  the STP distribution; {that is}, 
there is an  open interval %in $0\le \theta \le 2\pi$} 
on the unit circle where no {spiking time} point is detected. 
{ In \cite{PBKpre,BaptistaPRE08}, such a 
localization of {the} STPs (obtained from {a} sufficiently long time series) is 
mathematically described in the following {manner.}}   
{Let} the distribution of {the} STPs be included in a set $\mathcal{D}$ 
on the unit circle; $\mathcal{D}$ is localized if there exist  {  
 open sets $\Lambda_i$ on the circle}   
such that $\mathcal{D}\cap \sum_i\Lambda_i=\emptyset $.  
It should be noted that only one parameter $V_{th}$ 
is {required} to detect {the} CPS by this algorithm.

\section{Results}\label{result}
In this section, we mainly show the results for 
the periodic  driving force with the frequency at $f_0=0.9$ {Hz}. 
This value is  close to the natural frequency $f^*$ {of the fast mode}. 
In this case, {the} {quiescent periods} of $V$ disappear when {the} forcing amplitude $K$
 increases; {that is}, $V$ exhibits spiking without {quiescent periods}.  
  {Then, we find {the} CPS between {the} single spikes and the driving force.}  
 Moreover, 
we observe two characteristic phenomena around the transition 
point to {the} CPS.  {{One} {phenomenon shows} a power-law scaling against $K$, which is exhibited in the average time interval between 
successive phase slips. The scaling exponent} undergoes an unsmooth change 
as {$K$} is varied.   
{The} other {phenomenon shows} a stepwise behavior   
observed in Kuramoto's order parameter 
for the {driving force} strength before the transition {(shown in Fig. \ref{fig:worefracOP.eps})}.  

 In addition to the results of the case $f=f_0$, 
we also show brief results for the 
periodic driving force with frequency  
$f= f_1=0.92$ {Hz}. In this case, {the} 
{quiescent periods} of $V$ do not disappear {(shown in Fig. \ref{fig:PLK302O92.eps})}, even if the value of $K$ 
increases to its maximum value of $K\ {\approx} \ 0.5$ {for} the case $f=f_0$. 
As will be explained {later} in detail, a stepwise behavior  
can be observed as well {by using} another observation variable. 
All the characteristic phenomena are inherent 
to systems with multiple time scales. 
 
\subsection{Detection of Phase Synchronization}
\subsubsection*{Detection by {Spiking Time} Points} 

\begin{figure}[htbp]
   \begin{minipage}{.28\textwidth}
 \begin{center}
    \includegraphics[keepaspectratio=true,height=.13\textheight]{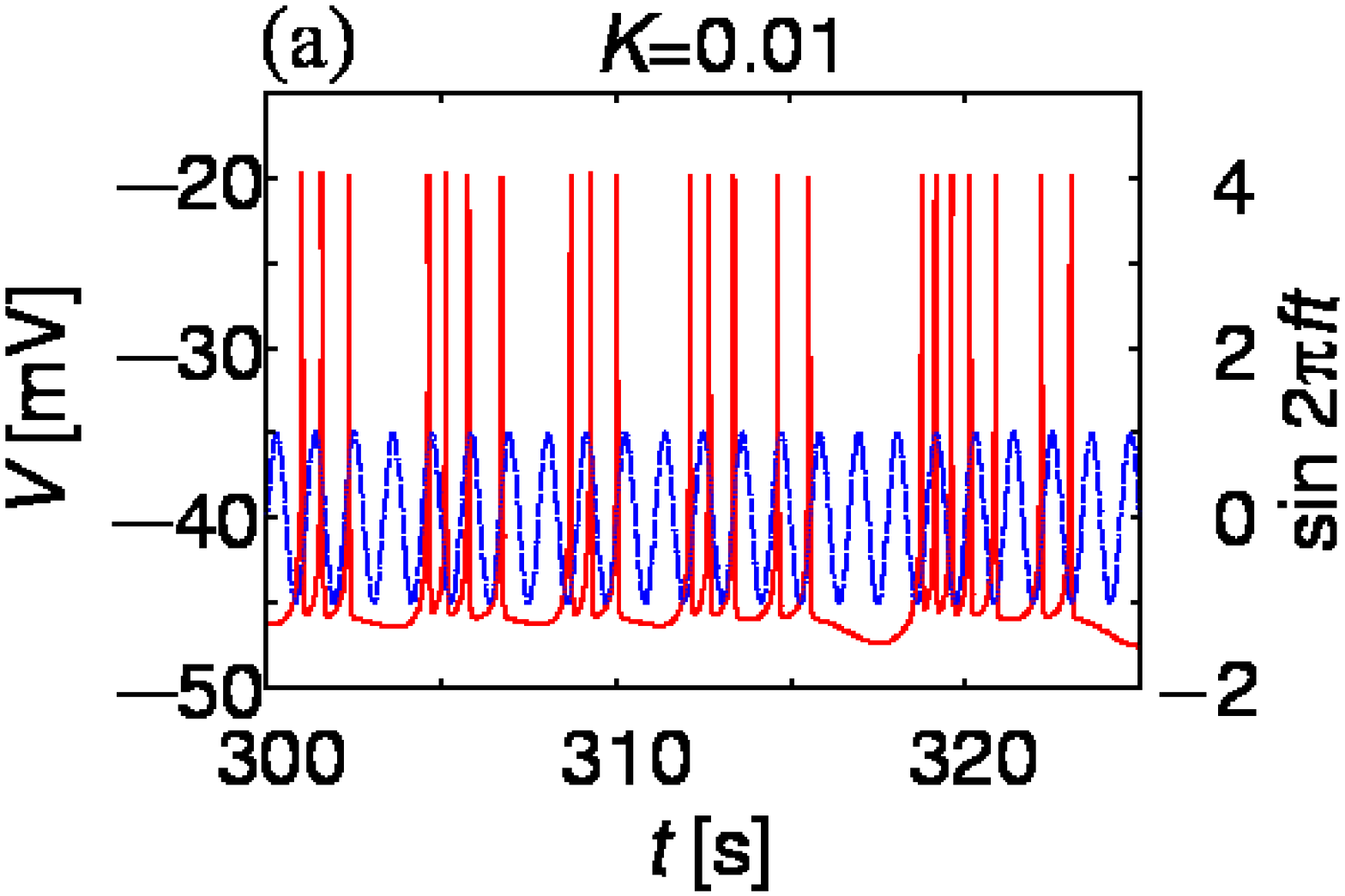}
  \end{center}
 \end{minipage}
  \begin{minipage}{.17\textwidth}
  \begin{center}
  \rotatebox{0}{\includegraphics[keepaspectratio=true,height=.13\textheight]{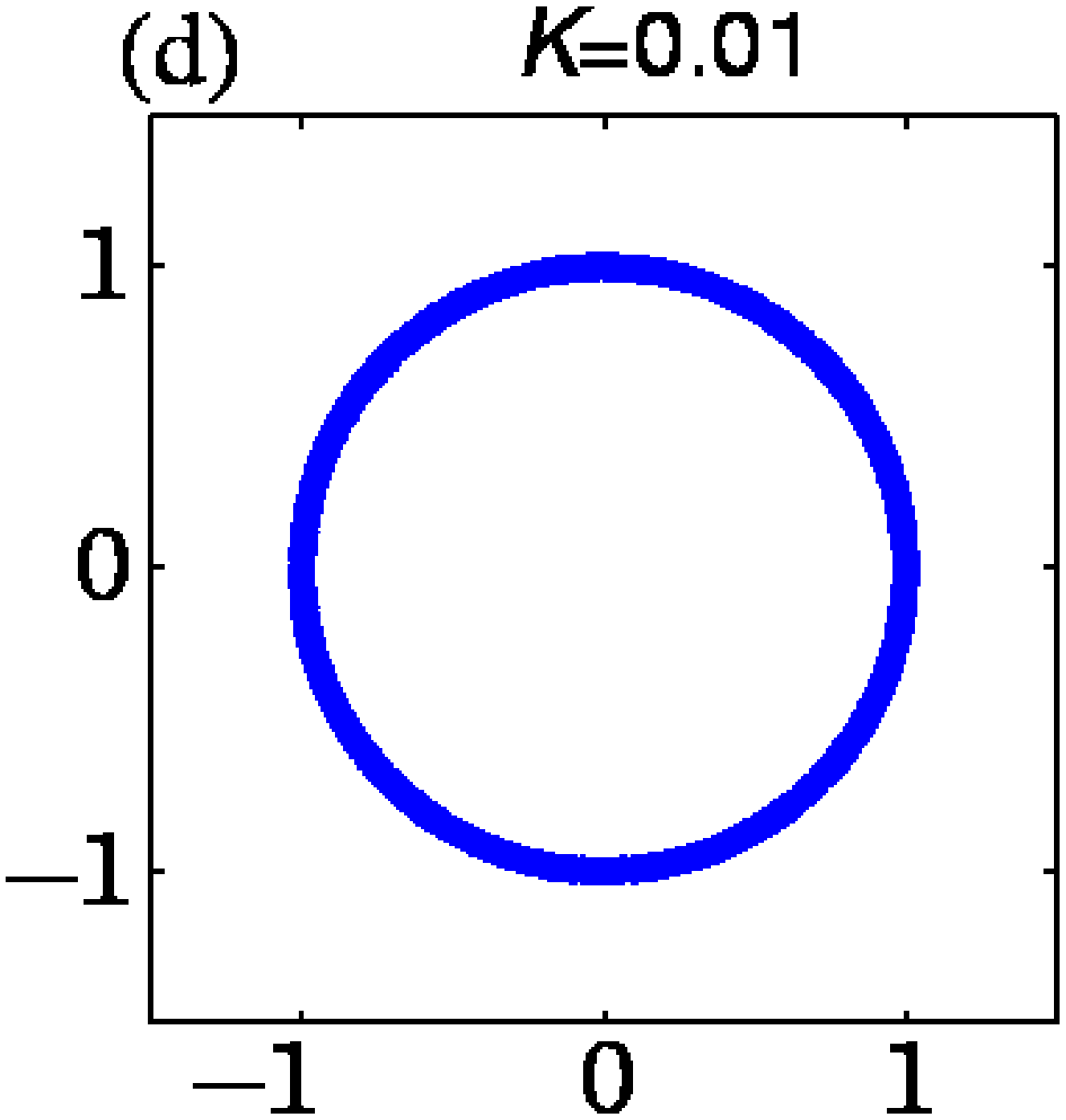}}
  \end{center}
  \end{minipage}
     \begin{minipage}{.28\textwidth}
  \begin{center}
\includegraphics[keepaspectratio=true,height=.13\textheight]{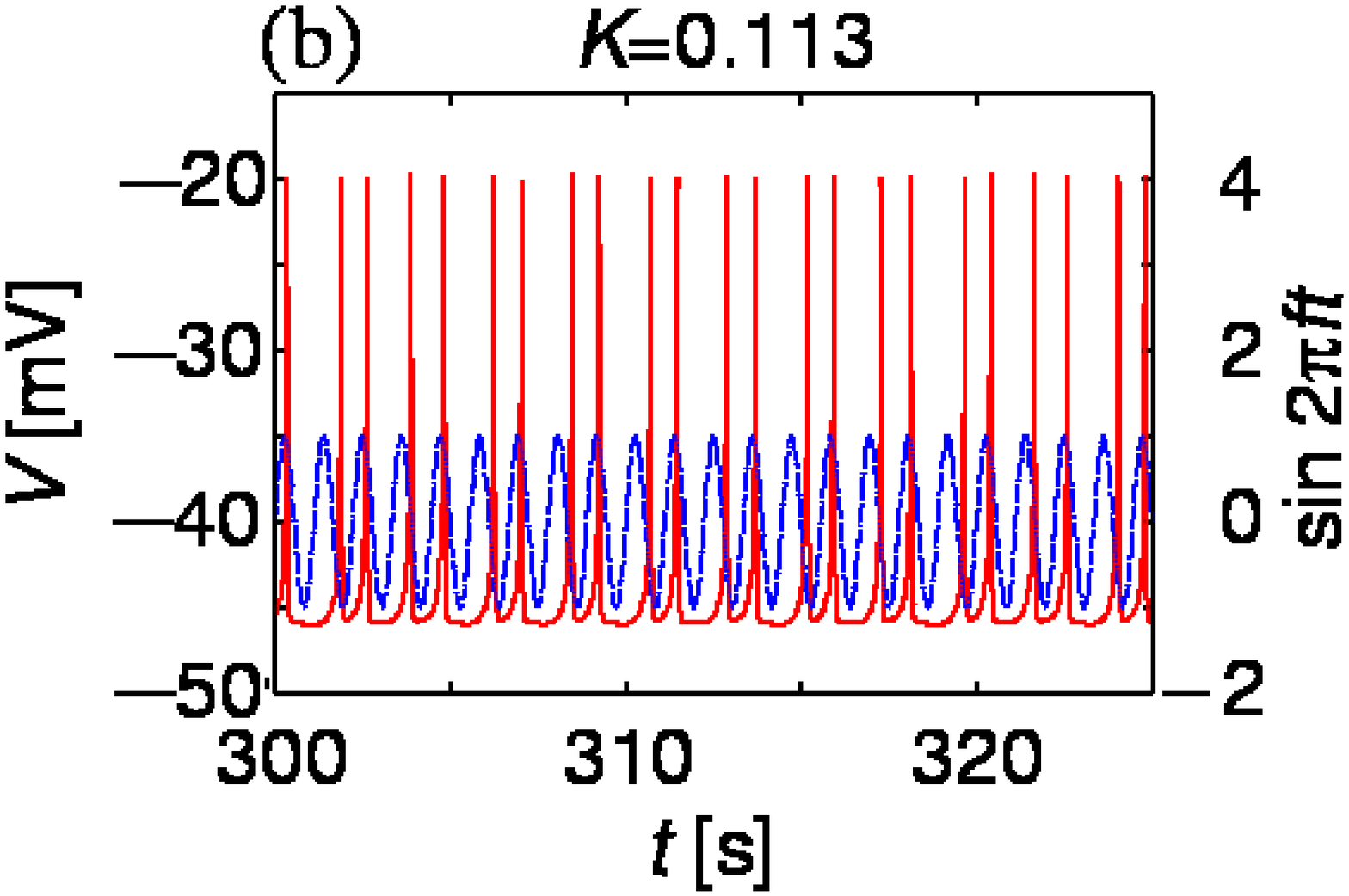}
  \end{center}
  \end{minipage}
      \begin{minipage}{.17\textwidth}
  \begin{center}
   \rotatebox{0}{\includegraphics[keepaspectratio=true,height=.13\textheight]{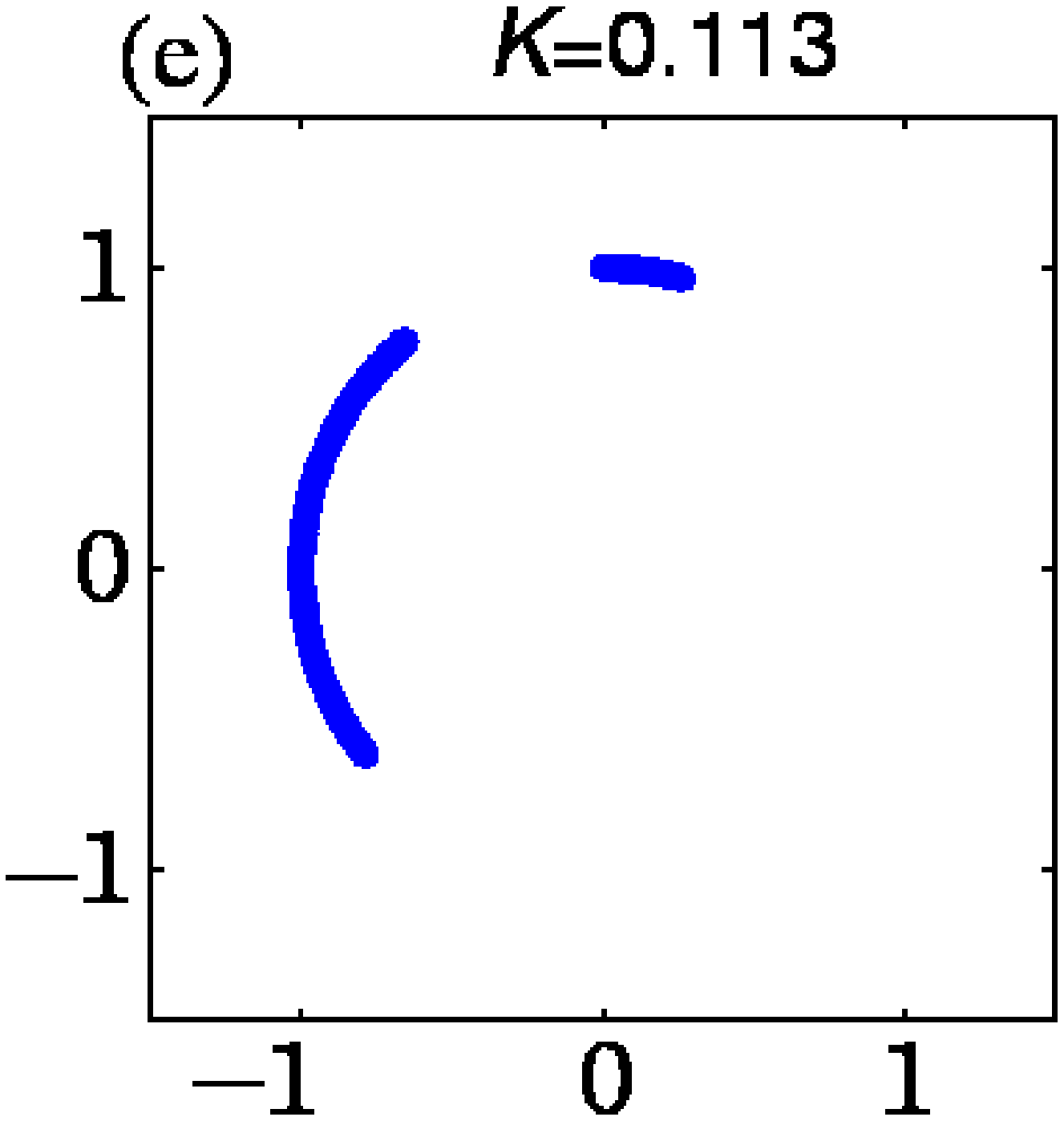}}
  \end{center}
  \end{minipage}
      \begin{minipage}{.28\textwidth}
  \begin{center}
      \includegraphics[keepaspectratio=true,height=.13\textheight]{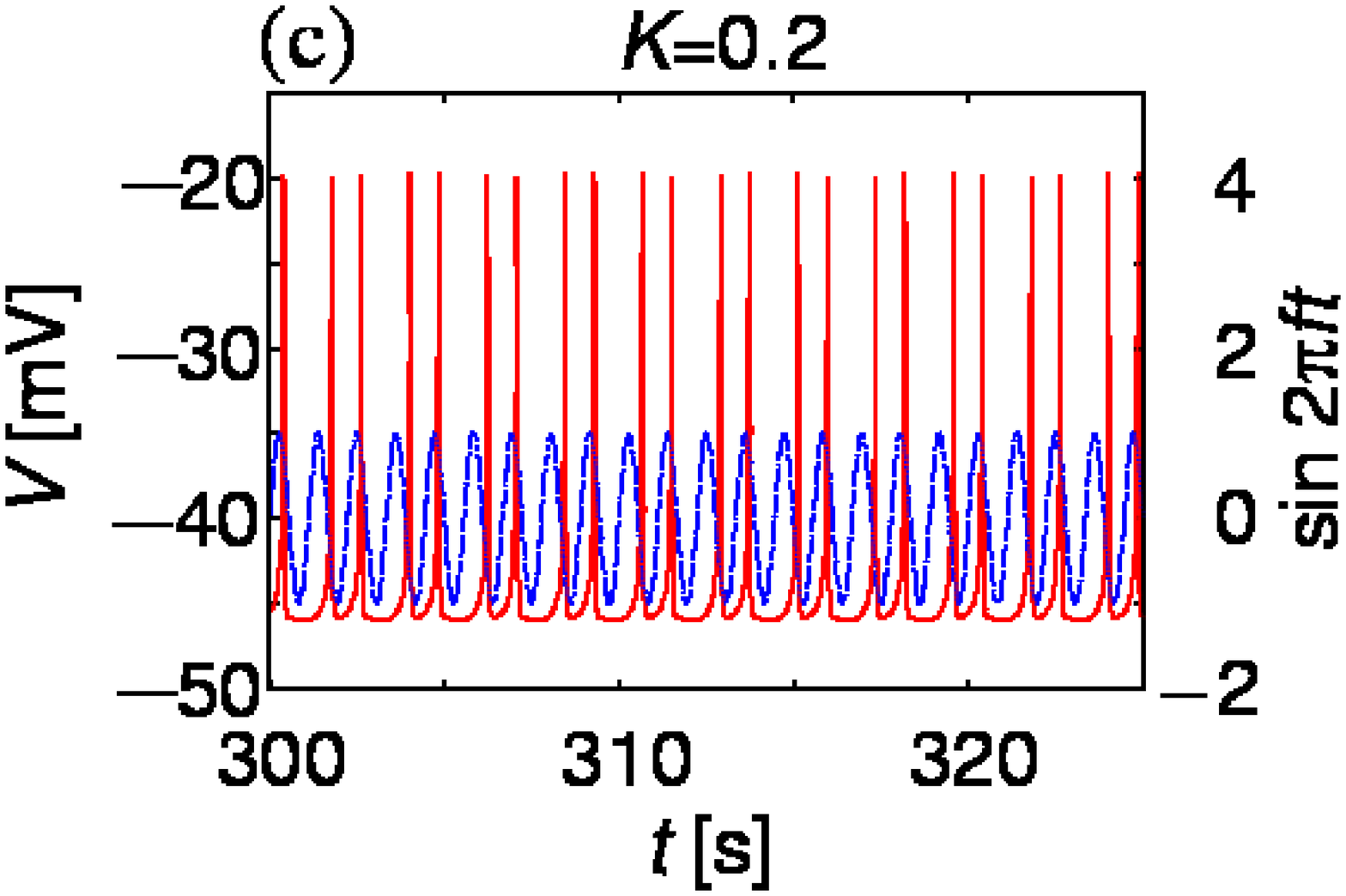}
  \end{center}
  \end{minipage}
    \begin{minipage}{.17\textwidth}
  \begin{center}
   \rotatebox{0}{\includegraphics[keepaspectratio=true,height=.13\textheight]{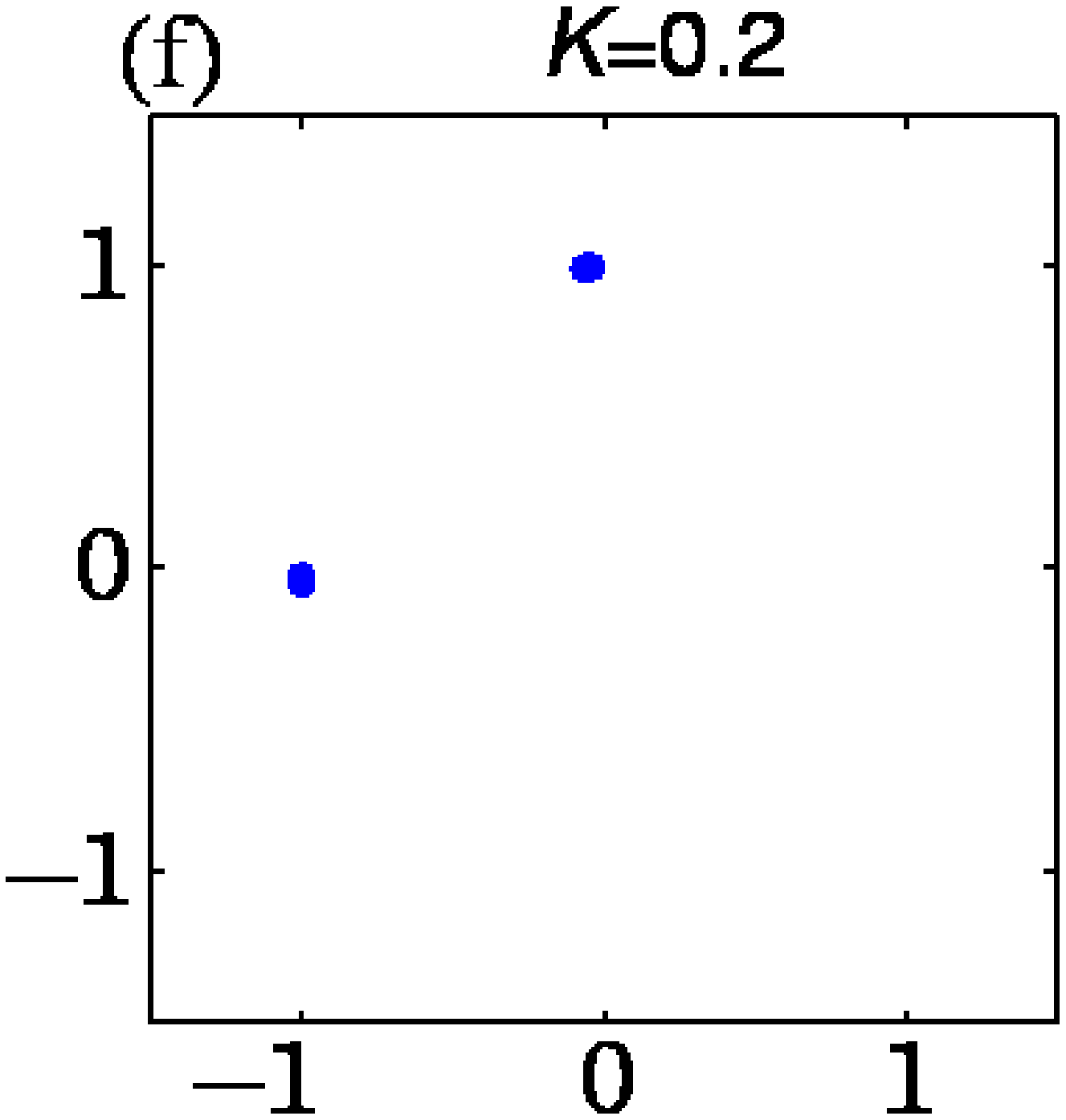}}
      \end{center}
  \end{minipage}
    \caption{(Color online) Time series of $V(t)$ (red solid line)  for driving forces (blue dotted line) with 
    different amplitudes: (a) $K=0.01$, (b) $K=0.113$, (c) $K=0.2$. {The corresponding} STPs appear on the unit circle in (d), (e), and (f),  respectively. {The length of {the} time series for plotting  
    {the} STPs is 1000 s.}}
  \label{fig:nonCPSK001O9.eps}
\end{figure}

Figures \ref{fig:nonCPSK001O9.eps}(a)--\ref{fig:nonCPSK001O9.eps}(c)  show 
the time series of $V(t)${,} together with the sinusoidal driving force. 
For {a weak driving force} with $K=0.01$, as shown in Fig. \ref{fig:nonCPSK001O9.eps}(a), 
there is no synchronization 
between the spikes and the force. On the other hand, 
when $K=0.113${,}  as shown in Fig. \ref{fig:nonCPSK001O9.eps}(b), 
 the system exhibits phase synchronization{,}  
i.e., {a} one-to-one correspondence between a single spike and 
one period of the driving force.  
Additionally, there are fluctuations in  the inter-spike intervals 
(ISIs). 
Therefore, {this state is considered to be one that exhibits CPS}. 
In the case of {a strong driving force} with $K=0.2$, as shown in Fig. \ref{fig:nonCPSK001O9.eps}(c),  
classic phase locking (CPL)  is observed with 
two periodically alternating ISIs. 
More precisely, each pair of successive  spikes 
{is observed} at the same points  in the period of the force.

 While the state {shown} in Fig. \ref{fig:nonCPSK001O9.eps}(a) {does not exhibit} phase 
 synchronization in terms of STPs,  
 the states {shown in Figs}. \ref{fig:nonCPSK001O9.eps}(b) and 3(c) {exhibit} phase synchronization.  
 Figures \ref{fig:nonCPSK001O9.eps}(d)--3(f) show {the} STPs 
 on the unit circle for a certain time interval. 
 {The length of {the} time series for plotting the STPs is 1000 s. This length of 
 {the} time series is sufficiently long to determine the localization of the distribution of {the} STPs{,} as mentioned in the definition of the localization of {the} STPs.}
 In Fig. \ref{fig:nonCPSK001O9.eps}(e), {when} $K=0.113$,  we can detect   
 {CPS} 
 between the spikes and the force, because {the} STPs are localized  
 yet distributed  on the  unit  circle. 
 {As shown in} Fig. \ref{fig:nonCPSK001O9.eps}(f), {when} $K=0.2$, 
 {the} CPL can be detected 
 {on the basis of} the fact that all {the} STPs are 
 concentrated at two points on the unit circle. 
 This means that each pair of successive spikes completely synchronizes 
 with two periods  of the force. 
 However,  no synchronization can be detected 
 in Fig. \ref{fig:nonCPSK001O9.eps}(d), {when} $K=0.01$,  {because 
 the STPs are not localized. In other words the entire} circle 
 is filled with STPs.

\begin{figure}[htbp]
  \begin{minipage}{.45\textwidth}
  \begin{center}
    \includegraphics[keepaspectratio=true,height=50mm]{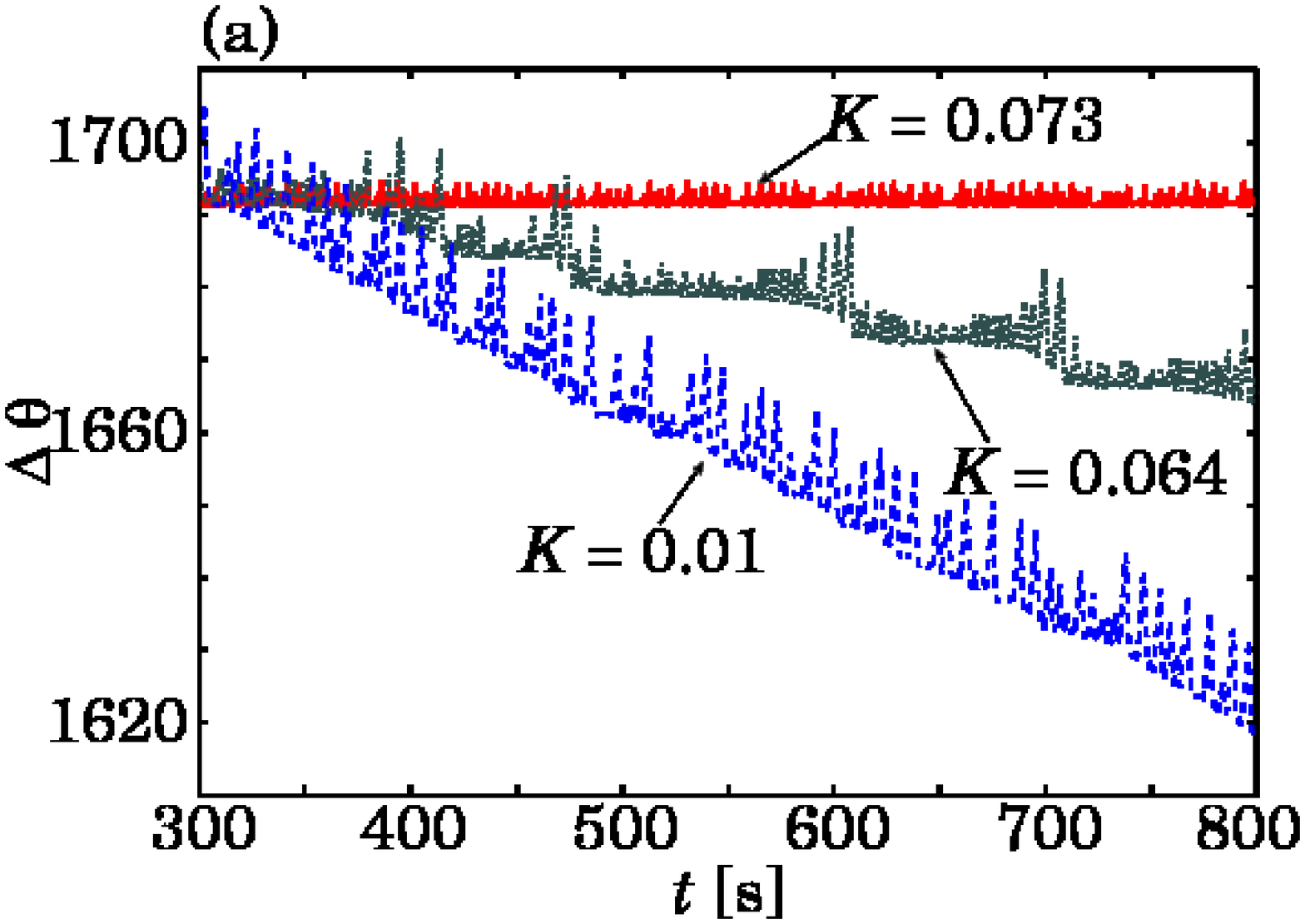}
  \end{center}
  \end{minipage}
  \begin{minipage}{.45\textwidth}
  \begin{center}
    \includegraphics[keepaspectratio=true,height=50mm]{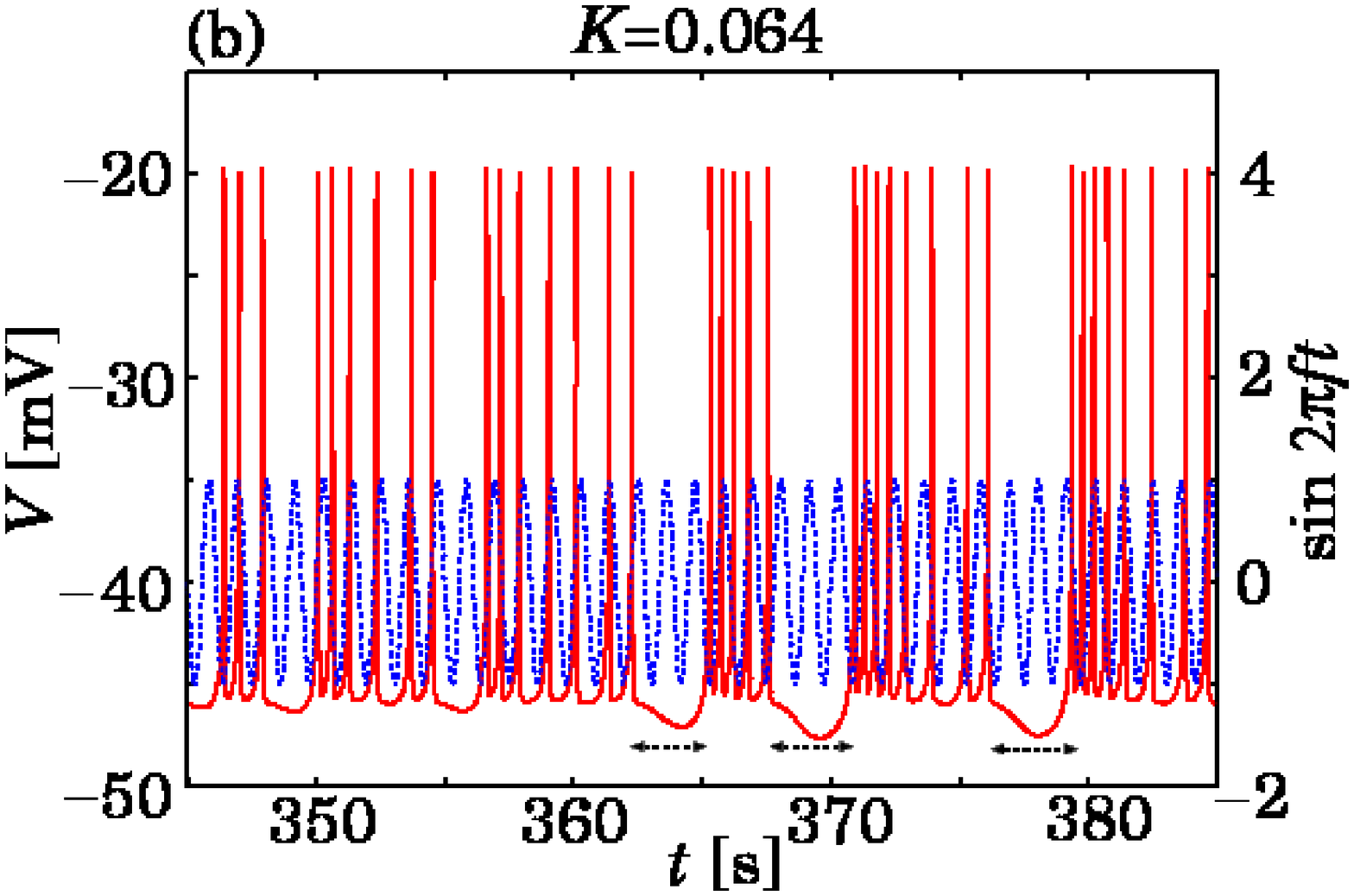}
  \end{center}
  \end{minipage}
    \caption{(Color online) (a) Time evolution of the phase difference 
    $\Delta \theta$ for $K=0.01$ (blue dashed line), $0.064$ 
    (green dotted line), 
    $0.073$ (red solid line). (b) Time series of $V(t)$ (red solid line)   with the periodic force 
    (blue dotted line) for $K=0.064$. Phase slips {are observed} in the region indicated 
    by the dashed arrows.}
  \label{fig:phaseslip.eps}
\end{figure}

\subsubsection*{Detection by Phase Difference} 

To confirm the {CPS,} as {shown} in 
Fig. \ref{fig:nonCPSK001O9.eps}(b){,} from another perspective, 
let us define the phase difference $\Delta \theta$ between 
the spiking oscillation and the driving force and then observe  
its time evolution for different {values of} $K$  around the transition point. 
Suppose that  {at each instance when the membrane potential exceeds the threshold value}, the phase variable of 
the spiking oscillation, $\theta_{\text{s}}$, increases by $2\pi$. 
The  instantaneous 
 phase variable of the external force, $\theta_{\text{e}}$, is determined 
 at the {spiking time}  without taking it {to be} modulo $2\pi$ { unlike $\theta_n$ considered in the case of STPs in Section \ref{model}}. 
Thus,  the phase difference is defined as $\Delta \theta=\theta_{\text{e}}-\theta_{\text{s}}$. 
 {It should be noted} that $l\ :\ m$ phase synchronization between  
spikes and the external force can be defined as 
$| m \theta_s- l \theta_e |<\text{const.}$

Figure \ref{fig:phaseslip.eps}(a) shows the time evolution of 
$\Delta \theta$ for $K=0.01$, $0.064$, and $0.073$. 
For $K=0.01$, the time evolution of $\Delta \theta$ 
shows an oscillation with {decreasing tendency}. 
For $K=0.064$, $\Delta \theta$ temporarily fluctuates 
within a bounded region (plateau) but sometimes exhibits a sudden phase slip.    
Finally, for $K=0.073$, $\Delta \theta$  {always} fluctuates 
within a bounded region; that is, phase slip {does not occur}. 
{It should be noted} that there {exists} a transition 
point to CPS near $K=0.073$.  Therefore, we can confirm that the state {shown} in 
Fig. \ref{fig:nonCPSK001O9.eps}(b) represents CPS  in the sense 
of the conventional definition{,} as well, {given that} $K=0.113$ is beyond 
the transition point. It should {also} be {noted} that phase slips occur  
when the {quiescent period} of $V$ takes  a relatively long time. 
These slips are indicated by the dashed 
arrows in Fig. \ref{fig:phaseslip.eps}(b).

%%scaling for phase slip has been moved to later

\begin{figure}[htbp]
  \begin{center}
    \includegraphics[keepaspectratio=true,height=45mm]{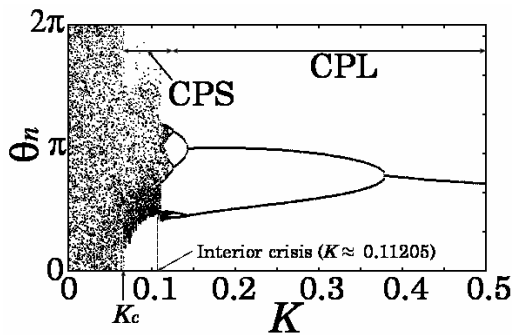}
     \end{center}
  \caption{Bifurcation diagram of $\theta_n$ with respect to $K$. 
  CPS and CPL represent chaotic phase synchronization and 
  classic phase locking, respectively.}
  \label{fig:worefractoriness.eps}
\end{figure}

\subsection{Route to Phase Synchronization and Phase Locking} 

Let us clarify the route to phase synchronization by 
the dynamics of $\theta_n$  for  the driving force at the $n$th spike{, which is an angle 
of {the} STPs on the unit circle}. 
The bifurcation diagram for $\theta_n$ with respect to $K$ is 
shown in Fig. \ref{fig:worefractoriness.eps}. 
For values of $K$ that are relatively small,  $\theta_n$ can take 
any value in the range between $0$ and $2\pi$, which means that {CPS does not occur} (a non-CPS state). 
After the first transition at $K=K_c\ {\approx} \ 0.073$, 
the value of $\theta_n$ is confined  within a localized region. 

\begin{figure}[t]
  \begin{center}
    \includegraphics[keepaspectratio=true,height=65mm]{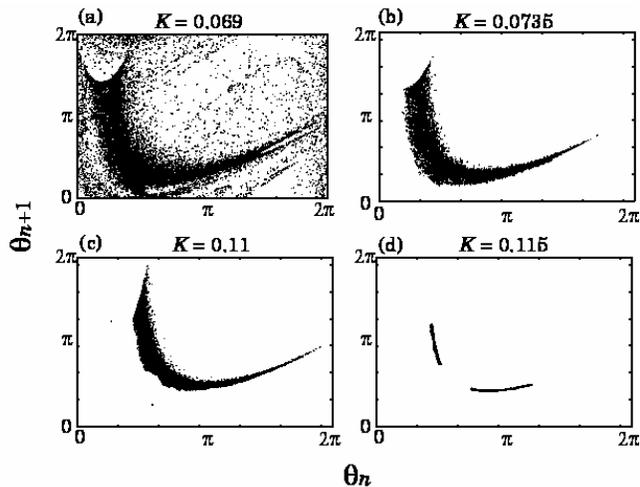}
  \end{center}
  \caption{Return plots  of $\theta_n$ in the space $(\theta_n,\ \theta_{n+1})$ for (a) $K=0.069$, (b) $K=0.0735$, (c) $K=0.11$, and (d) $K=0.115$.}
  \label{fig:criseso9.eps}
\end{figure}

First, we examine the system behavior around $K=K_c$.  
Figure \ref{fig:criseso9.eps} shows the  return plots  in the 
space $(\theta_n,\ \theta_{n+1})$. The points {shown} in Fig. \ref{fig:criseso9.eps}(b) ($K=0.0735$)  
are distributed within a limited region, {whereas} the points {shown} in Fig. \ref{fig:criseso9.eps}(a) 
($K=0.069$) 
 are distributed throughout the entire space. 
 Therefore, these plots imply that the system undergoes 
 a boundary crisis \cite{OTT}.  
 Here, the crowding of points {shown} in Fig. \ref{fig:criseso9.eps}(b) represents
{CPS} (a CPS state) {as in the case of} 
the localization of {the} STPs on the unit circle, {implying} that phase slip {does not occur}.

Moreover, the second transition at $K\ {\approx} \ 0.11205$ 
seems {to be} an interior crisis \cite{OTT}, which is {indicated by the change}  
between Fig. \ref{fig:criseso9.eps}(c) ($K=0.11$)  
after the crisis and Fig. \ref{fig:criseso9.eps}(d) ($K=0.115$)  before the crisis. 
More precisely, the two disconnected attracting sets {shown} in Fig. \ref{fig:criseso9.eps}(d) 
are included in the single attractor {shown} in Fig. \ref{fig:criseso9.eps}(c).  
After the second transition, a typical sequence of inverse period-doubling 
bifurcations  occurs. As a result, {CPL} (a CPL state) 
is observed in the 
regime where the driven system fires periodically.

\begin{figure}[htbp]
  % \begin{minipage}{.45\textwidth}
     \begin{center}
    \includegraphics[keepaspectratio=true,height=50mm]{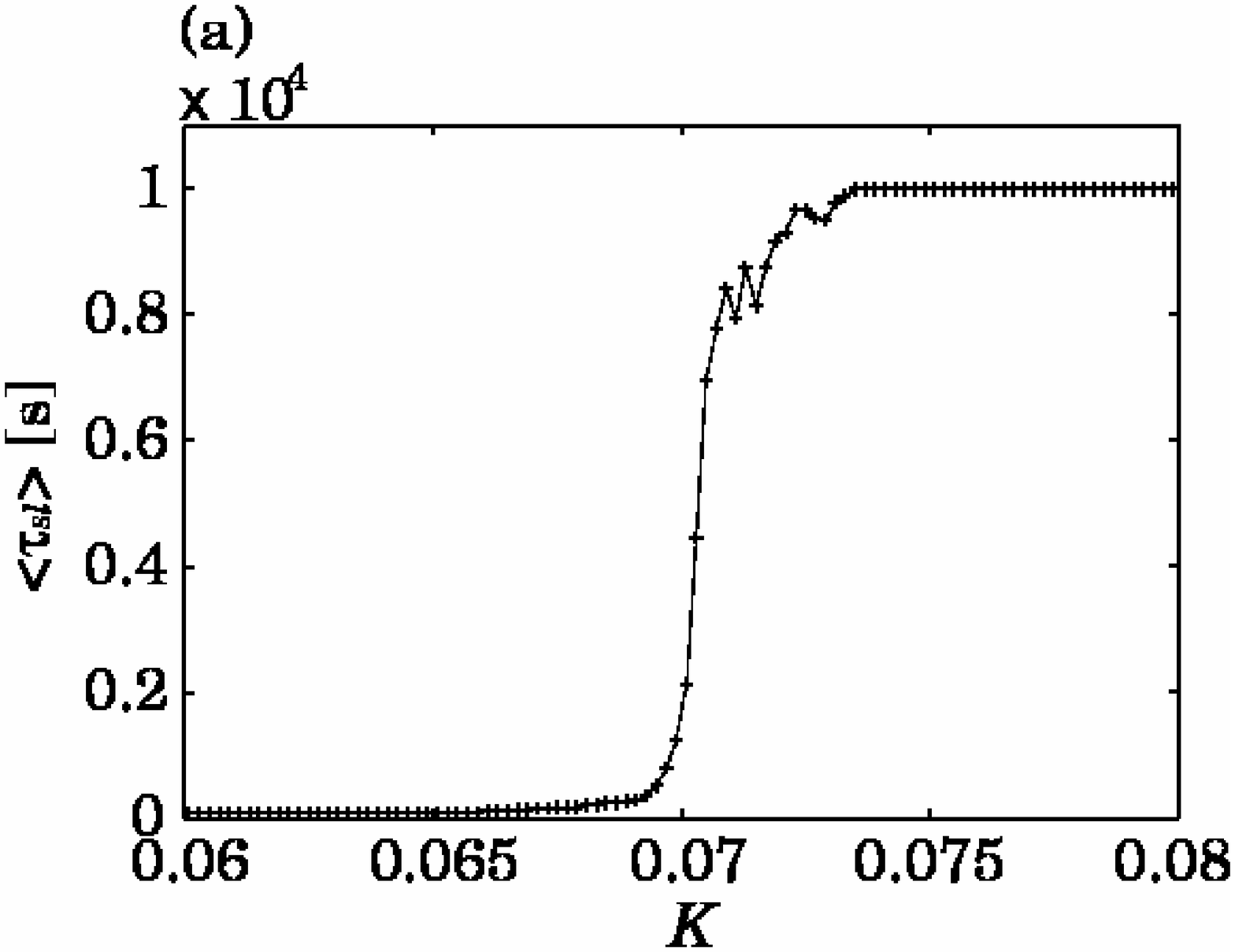}
  \end{center}
  %\end{minipage}
   %\begin{minipage}{.45\textwidth}
  \begin{center}
    \includegraphics[keepaspectratio=true,height=50mm]{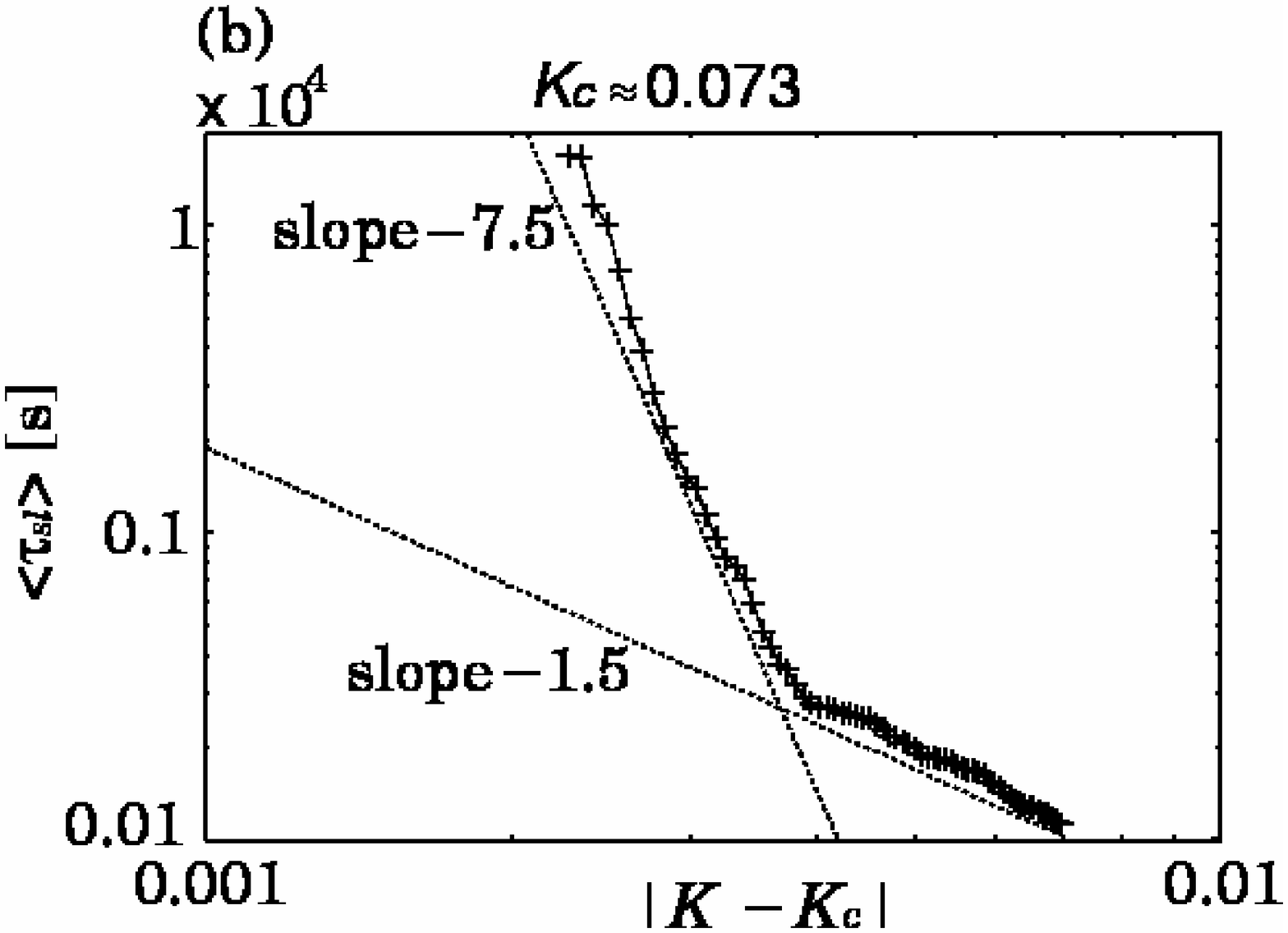}
  \end{center}
  %\end{minipage}
  \caption{(a) Average time intervals between  phase slips 
  with respect to $K$ around the phase transition point. 
  (b) Log-log plot of the average time intervals with respect to 
  the difference between {the} parameter $K$ and its critical  
  value $K_c\ {\approx}\  0.073$.  The slopes represent {the} scaling exponents. { The results 
  are averaged over 100 and 1000 different realizations for (a) and (b), respectively.}}
  \label{fig:sliploglog.eps}
\end{figure}

\subsection{Characteristics around the Transition Point}
%%moved from the section of phase slips
\subsubsection*{Scaling Behavior for Phase Slips} 
We characterize the average time interval 
$\langle \tau_{sl} \rangle$ between two successive  phase slips  
(i.e., the plateau length) with respect to 
$K$ around the transition point, as shown in Fig. \ref{fig:sliploglog.eps}(a) \cite{realization}.   
Figure \ref{fig:sliploglog.eps}(b) shows the log-log plot of 
$\langle \tau_{sl}\rangle$ {in dependence on} $|K-K_c|$, where $K_c\ {\approx}\  0.073$ 
is the transition point to CPS. Here,  $K_c$ is 
approximately determined  by the point at which 
the distribution of {the} STPs 
begins to become localized.  
We  numerically  find a power-law 
scaling   $\langle \tau_{sl}\rangle \sim |K-K_c|^{\gamma}$,  
where the scaling constant suddenly changes from $\gamma \ {\approx}\  -1.5$ 
to {$\gamma\  {\approx}\  -7.5$}.  Although the range of 
the scaling region is relatively short,  
this scaling behavior is clearly 
different from that observed for a system with a 
single time scale. The scaling behavior with a single time scale is related 
to type-I intermittency and is described by 
$\log \langle \tau_{sl}\rangle \sim |K-K_c|^{-1/2}$ \cite{PikovskyChaos97,footnote}.

In general,  { for a periodically driven chaotic system with a  
single time scale and a single rotation center, 
a simplified mapping model 
can explain {the} transition to {CPS} between 
the system and {the} driving force. In the map model, 
the {boundary} between {the} synchronization state and 
{the} non-synchronization state is explained by a saddle-node bifurcation of  
unstable and stable periodic orbits of the map
\cite{PikovskyChaos97}.}  However, 
in the present system,  
phase locking cannot  
simply be related to  a saddle-node bifurcation,  
 {because} multiple time scales {exist}.  
 That is, {the spiking period is characterized by fast oscillations},  related to 
 {the} variables $V$ and ${q}$,  {and the} {quiescent period} {is characterized by slow oscillations},  related to  {the} variable $C$. 
Hence,   
{the dynamics of the system on the threshold, 
which corresponds to the map model,}    
  is affected by both the fast and {the} slow oscillations.  {{Therefore}, the  
  mechanism of {a} sudden change in the scaling law {differs} from the 
  case of a single time scale but is still an open problem.}   
%%%%%%%%

\begin{figure}[htbp]
\begin{minipage}{.45\textwidth}
  \begin{center}
    \includegraphics[keepaspectratio=true,height=45mm]{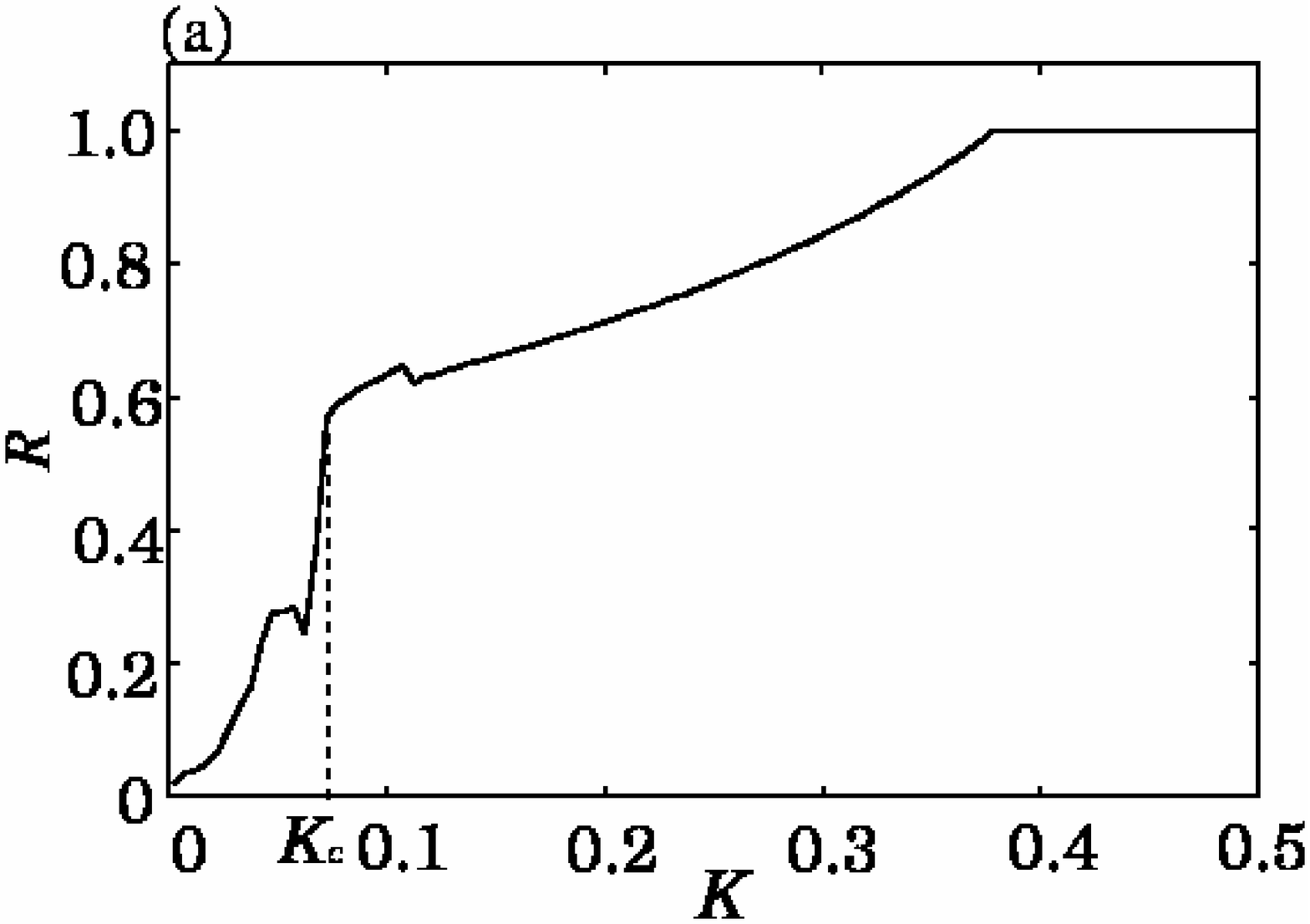}
    \end{center}
    \end{minipage}
    \begin{minipage}{.45\textwidth}
  \begin{center}
    \includegraphics[keepaspectratio=true,height=45mm]{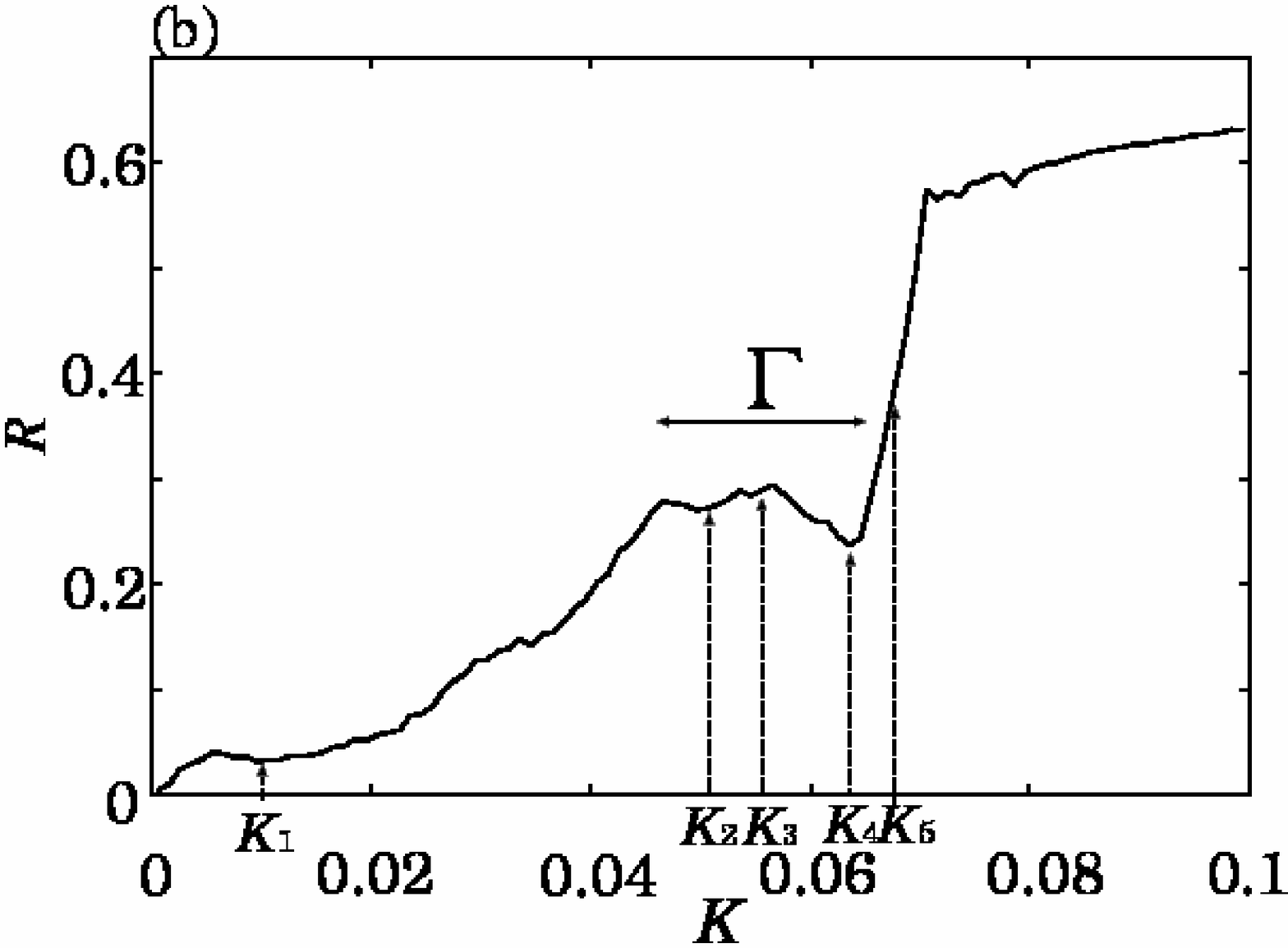}
    \end{center}
    \end{minipage}
  \caption{(a) Order parameter $R$ as a function of $K$. 
  The dashed line indicates $K_c$ or  the transition point  to CPS. 
  (b) Enlargement of (a) for $0\le K \le 0.1$.  {It should be noted} that $\Gamma$ 
  {denotes} the step region before the transition to CPS. { The 
  results are averaged over 1000 different realizations.}} 
  \label{fig:worefracOP.eps}
\end{figure}

\subsubsection*{Stepwise Behavior Measured by Order Parameter} 
The distribution of {the} STPs is now used to detect phase synchronization.  
In order to quantify the degree of phase synchronization,  
we employ Kuramoto's order parameter defined as   
$R=|\sum^N_{j=1} \exp(i \theta_j)/N|$, where $N$ is 
the number { of STPs} \cite{Kuramoto}. 
{The parameter} $R$ satisfies $0\le R \le 1$. This corresponds to 
the amplitude of {an average} of STPs. 
The more localized the distribution of the STPs {is}, the greater {is} the value of $R$. 

Figure \ref{fig:worefracOP.eps}(a) shows the order parameter 
$R$ as a function of $K$, with  
$R$ averaged over 1000 {different realizations} for a time interval of 1000 s. 
The transition point $K_c$  to CPS is denoted 
by the dashed line in Fig. \ref{fig:worefracOP.eps}(a). 
We  find that  a stepwise behavior precedes the transition  to CPS,  
with the step region $\Gamma${,}  
as shown in  Fig. \ref{fig:worefracOP.eps}(b). 
This stepwise behavior {indicates} that 
there exists a region 
{between the non-CPS state and the CPS state,}  
where the degree of synchronization is not sensitive to $K$. 
In addition, $R$ decreases just before the transition point.  
To the best of our knowledge, such a stepwise behavior in 
{the}  transition has not been observed 
for CPS in coupled systems with a single time scale. 
 In what follows, we will explain how the stepwise transition 
and the decrease in $R$ are related to the existence of both the 
slow and {the} fast dynamics in the present system.   
{In particular}, we will investigate the probability density 
distribution of $\theta_n$ at $K=K_i,$ with $i=1,\ldots,5$, as  
indicated in Fig. \ref{fig:worefracOP.eps}(b).

\begin{figure}[t]
  \begin{center}
    \includegraphics[keepaspectratio=true,height=45mm]{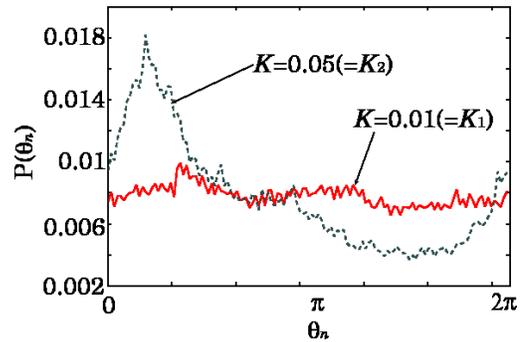}
  \end{center}
  \caption{(Color online) Probability density distributions of $\theta_n$ for $K=0.01$ (red solid line)  and ${K=}\ 0.05$ (green dashed line).}  
  \label{fig:P_theta_o9_2.eps}
\end{figure}

Figure \ref{fig:P_theta_o9_2.eps} shows the shape of 
the probability density distributions $P(\theta_n)$  of {the} STPs 
on the unit circle{, i.e.,} the distributions of  $\theta_n$  
for $K=K_2=0.05$ and for $K=K_1=0.01$. 
 {It should be noted}  that CPS is not detected in either case; {that is}, 
 the points are well distributed around the circle. 
 However, we can find a peak in the probability 
 density distribution for   
$K=K_2$, {whereas}  the distribution for $K=K_1$ is almost uniform.   

 The appearance of the peak{,} as shown in Fig. \ref{fig:P_theta_o9_2.eps}{,}  
and a shift {in} the peak in the step region $\Gamma$, as explained in Appendix \ref{mech},  
are two consecutive stages that constitute the {entire} 
stepwise phenomenon. 
 In the first stage, with {an} increase {in} $K$ from zero  
 to a value near $K_2$, 
a peak  in $P(\theta_n)$ of the STPs emerges near $\theta_n=0.5$. 
Then, in the second stage, the position of 
the peak  shifts {because of the} effect of slow dynamics{, i.e.},   
 an increase {in} the number of short inter-burst intervals.   
During the shift of the peak, $R$ does not change very {significantly because}  the value of 
$K$ varies in $\Gamma$. Thus, the step region can be observed.  
These two stages are investigated in detail in Appendix \ref{mech}.

\subsection{CPS with {Quiescent Period}}

In the above results, we have {primarily} investigated {the}   
$1\ :\ 1$ phase synchronization between  
{the} spikes and each period of the driving force,  
wherein {quiescent periods} disappear for the values of $K$  after 
phase synchronization. On the other hand,  
when the frequency of the driving force is $f=f_1=0.92$ { Hz}, 
the {quiescent periods} do not disappear even if {the} amplitude $K$ increases 
to the same level as in the case $f=f_0=0.9$ { Hz}. 
{It should be noted} that {the} $1\ :\ 1$ {CPL} can be observed 
{for a}  
sufficiently large $K$, {e.g.,} $K=2$. For values of $K\ {<}\ 0.5$, 
  $1:1$ phase synchronization cannot be 
clearly observed between {the} spikes and the driving force. 
However, {if we define $\Theta_n$ 
 at the time when $V$ approaches  
the minimum voltage in each specific quiescent period, 
where $V$ decreases under the threshold 
$V^*=-47$,} we can detect CPS in the sense of localization 
of the distribution of {$\Theta_n$ on the unit circle}. 
{{It is important to note} that the change {in} the order parameter (derived from $\Theta_n$) 
with respect to $K$ does not depend on the value of $V^*${,}  
sensitively. {In other words}, the value of $V^*$ {that is} less than {approximately} $-46.4$ {causes}     
a stepwise transition in the order parameter for $f_0=0.92$ Hz using $\Theta_n${,}  
as shown below.}

\begin{figure}[t]
  \begin{center}
    \rotatebox{-90}{\includegraphics[keepaspectratio=true,height=70mm]{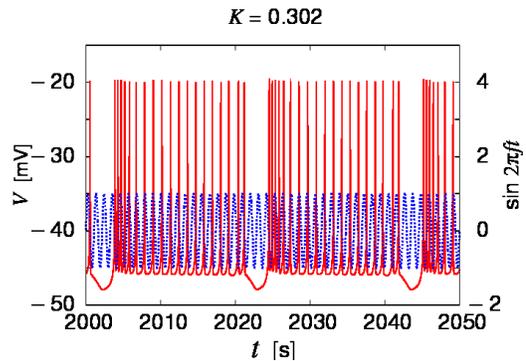}}
  \end{center}
  \caption{(Color online) $1\ :\ 19$ {CPL} between bursting (red solid  line)  and the periodic force (blue dotted line).}
  \label{fig:PLK302O92.eps}
\end{figure}

Figure \ref{fig:wrefractoriness.eps}(a) shows the bifurcation 
diagram of $\Theta_n$ {depending} on $K$.  
We also observe a stepwise behavior before a transition to 
{the} CPS in the variation of the order parameter 
computed from $\Theta_n$, as shown in 
Fig. \ref{fig:wrefractoriness.eps}(b). Moreover, 
in the region of CPS, fine tuning of $K$ {yields} 
$1\ :\ l$ {CPL} between a burst and $l$ periods 
of the force{, where}  $l=14,15,18,19$, etc. 
Figure \ref{fig:PLK302O92.eps} shows $1\ :\ 19$ {CPL} for $K=0.302$.

\begin{figure}[t]
  \begin{center}
    \includegraphics[keepaspectratio=true,height=45mm]{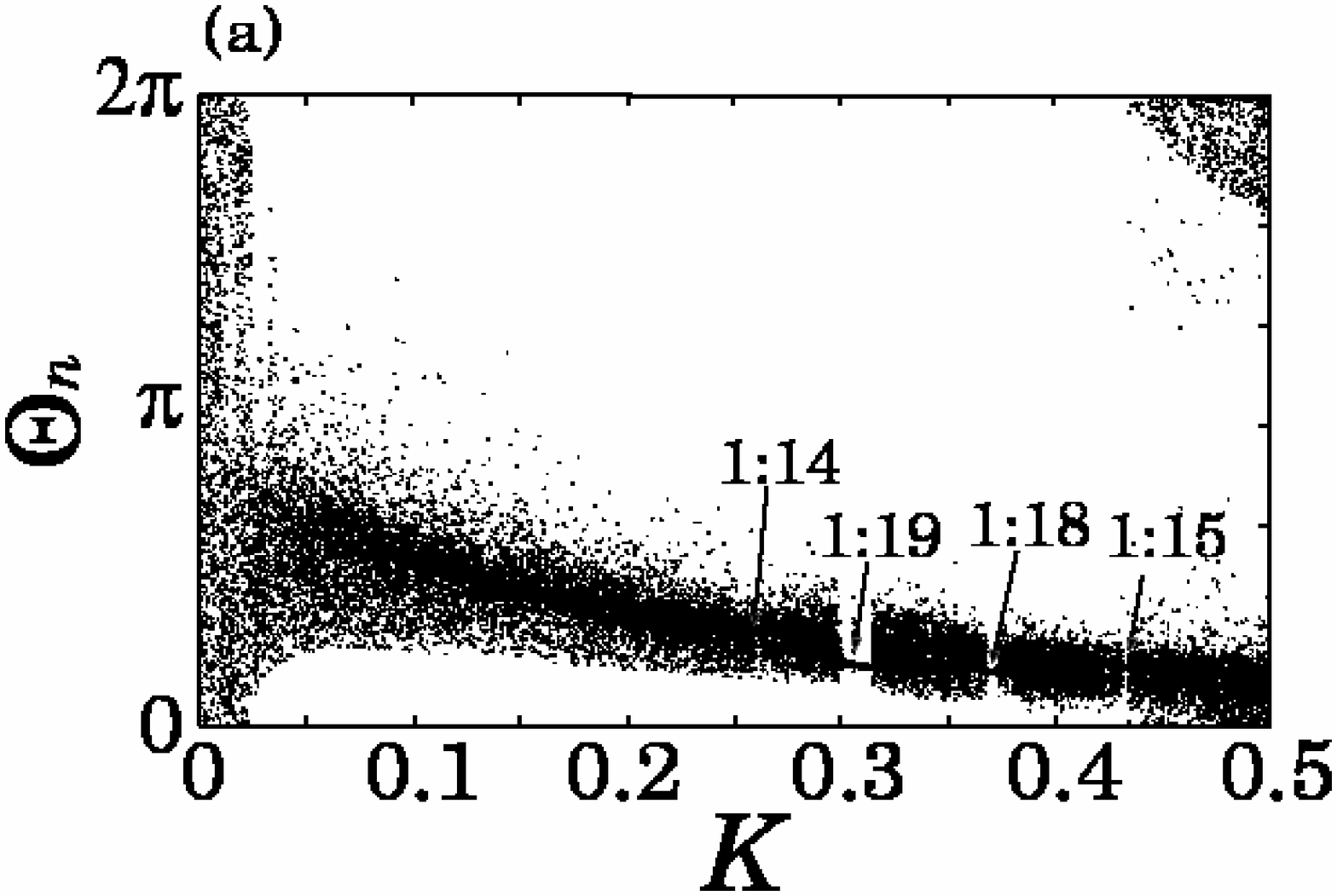}
  \end{center}
  \begin{center}
    \includegraphics[keepaspectratio=true,height=45mm]{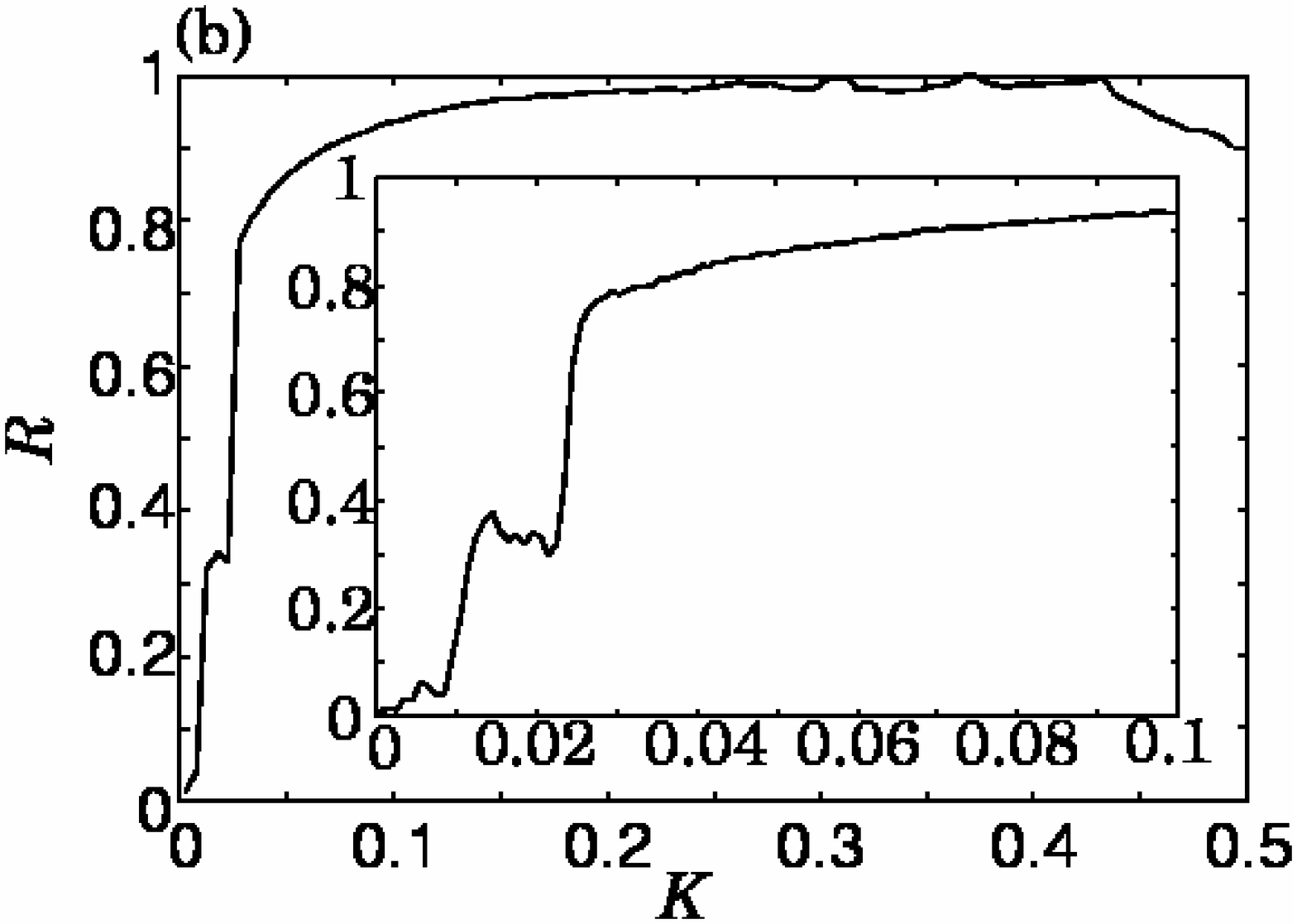}
  \end{center}
  \caption{(a) Bifurcation diagram of $\Theta_n$ with respect to $K$. (b) The corresponding values of {the} order parameter $R$. {The inset} {shows} a {magnified image} for $0\le K\le 0.1$. { The results are 
  averaged over 1000 different realizations.}} 
  \label{fig:wrefractoriness.eps}
\end{figure}

\section{Summary and Discussion}\label{summary}

We have investigated {CPS} 
in a spiking-bursting neuron model under periodic forcing  
with a small amplitude but with {a}  frequency as high as that of 
{the} spikes. 
First, we {observed} $1\ :\ 1$ {CPS} 
between {the} spikes and the periodic force. 
{This {CPS} has been 
detected {on the basis of the fact that} a set of points, {which are} conditioned by the phases of 
the periodic force at each {spiking time}, was concentrated on a 
sector of the unit circle.}

In addition to {CPS}, 
we {observed} two characteristic phenomena around 
the transition point to {CPS}. 
 {One phenomenon involves a} change in 
{the} power-law scaling  for {the} average time intervals 
between phase slips, as shown in Fig. \ref{fig:sliploglog.eps}(b).  
This scaling behavior is different from that {exhibited by} the conventional 
system, 
{ i.e., a chaotic system whose attractor has a single rotation 
center with only one characteristic time scale. In such a conventional 
system, the scaling exponent for the transition 
to {CPS} takes {a} unique value of $-1/2$.}  This might be because 
the phase synchronization in question cannot be simply 
associated with a saddle-node bifurcation, {owing} to 
the interaction between {the} slow and {the} fast dynamics. 
The {other phenomenon shows} a stepwise behavior before the transition 
to {CPS}, as shown 
in Fig. \ref{fig:worefracOP.eps}(b). This  {phenomenon} 
has been found by the observation that the degree of phase synchronization 
is not sensitive to the amplitude 
of the force just before the transition point. 
Moreover, we  found that {a decrease} in the degree of synchronization  
appears (at $K=K_4$ in Fig. \ref{fig:worefracOP.eps}(b)) 
even if the amplitude of the force is increased. 
The stepwise behavior and {this decrease} could be induced by 
the effect of slow dynamics ({see} Appendix \ref{mech}).

From the viewpoint of neuroscience,  our {system} might be regarded as
a simple model {whose fast driving force corresponds to sharp wave-ripples}. This {phenomenon involves} a very fast
oscillation of local field potentials observed in the hippocampus in the brain \cite{BuzsakiBook06}.  Let us discuss{,} below{,} a possible interpretation of our results in terms of {synchronization phenomena in} real neuronal systems. 

First, the observed phenomena in our model, {particularly CPS} 
with {a} fast driving force, can be interpreted as a consequence of the interaction between the ripples and the activity of a single neuron. In fact, it has been experimentally shown that such ripples synchronize in phase with {the} spikes of a single neuron \cite{Ylinen95,CsicsvariJNS99,KlausbergerNatN04}.

Furthermore, it has been suggested that firing sequences accompanying ripples in the hippocampal network form a representation of stored information. The replay of the firing sequences 
during sleep 
mediates a consolidation of memory for the stored information in the hippocampus and the neocortex \cite{Buzsaki96,SirotaPNAS03,GirardeauNatN09,DiekelmannNatRevN10}. In addition, some experimental results have indicated that such a memory replay is conducted on a shorter  time scale than {the} actual experience,   
where the spatiotemporal structure of the firing sequences plays a key role in the stored  information \cite{SkaggsSci96,NadsdyJNS99, LeeNeuron02,JiNatN06}.

Thus, if our spiking-bursting system with {a} fast driving force   
{can} be regarded as a model for such biological neurons, our result would imply that the precision of the temporal structure of {the} spiking patterns {might be} enhanced by {CPS} in real neuronal systems. {It should be noticed that CPS is not affected even if there exist small fluctuations in ISIs and that  
CPS extends the detection of the temporal structures of the 
firing patterns in the short time scale of a spiking-bursting behavior.}

Additionally, from another point of view, the slow oscillations along the lower envelope of the membrane potential $V$ {can}
 be considered as transitions between {the} UP and DOWN states in a cortical neuron. 
 {Specifically, we focus on the UP and DOWN states during the slow-wave sleep}.  
 {Here, {the} neurons in {the} UP state fire synchronously with higher frequency, {whereas} the activity of {the} neurons in {the} DOWN state is relatively quiescent \cite{Shu}.} 
 From this perspective, the disappearance of {the} {quiescent periods} (DOWN states) in the forced spiking in our simulation may be interpreted as a persistently depolarized UP state observed for cortical 
 neurons in {the} awake states.  
 In fact, the prolonged DOWN states during sleep  
 are likely to occur {owing} to a decrease {in the} excitatory input \cite{SteriadeJNS01}.
 Therefore, if this input {can} be regarded as our sinusoidal driving force, {CPS} may efficiently help a weak input to depolarize a DOWN state into a persistent UP {one}.  
 {Recently, Ngo {\em et al.} reported that a neural network model can {qualitatively} reproduce the experimental results in \cite{Shu} using a time-discrete map model{,} which is simpler 
 than {our model} \cite{Ngo}. }

{Therefore}, we may infer that our results on {CPS} at high frequency in the simple spiking-busting neuron model provide  
some suggestions for neuroscience to understand {the} mechanisms of the 
abovementioned  real neuronal activity, particularly in terms of nonlinear dynamics.

{The} following topics might be considered {from the viewpoint of 
extending this study}. 
First, {for the future study} {it would be important} to confirm whether 
our {findings} can be observed in other 
slow-fast models such as the Hindmarsh-Rose model \cite{HR} 
and {other} biophysical models \cite{FalckeBC00,VaronaNN01,PintoPRE00}.   
 {A} comparison with {the} other models will provide 
further insight into slow-fast dynamics in {the} neuronal spiking-bursting activity. 
Moreover, it should be  important to clarify the general onset mechanism 
of the observed phenomena  using map-based models, as was {clarified} in \cite{PikovskyChaos97, Yamada_etal}. 
{It should also be important} to investigate the response of coupled bursting systems to sinusoidal forcing in terms of the interaction between {the} slow and {the} fast dynamics. 

\begin{acknowledgements}
The authors are grateful to Prof. M. Tatsuno, Prof. H. Hata, Prof. M. Baptista, Dr. K. Morita, Dr. S. Kang, and Dr. G. Tanaka for their valuable suggestions. This work is partly supported by Aihara Complexity Modelling Project, ERATO, JST, and the Ministry of Education, Science, Sports and Culture, Grant-in-Aid for Scientific Research No. 21800089 and No. 20246026, and 
the Aihara Project, the FIRST program from JSPS, initiated by CSTP, and grants of the German Research Foundation (DFG) in the Research Group FOR 868 Computational
Modeling of Behavioral, Cognitive, and Neural Dynamics. 
\end{acknowledgements}

\appendix
\section{Mechanism of Stepwise Behavior}\label{mech} 
%%%Appendix%%%

\begin{figure}[t]
 \begin{minipage}{.45\textwidth}
  \begin{center}
    \includegraphics[keepaspectratio=true,height=45mm]{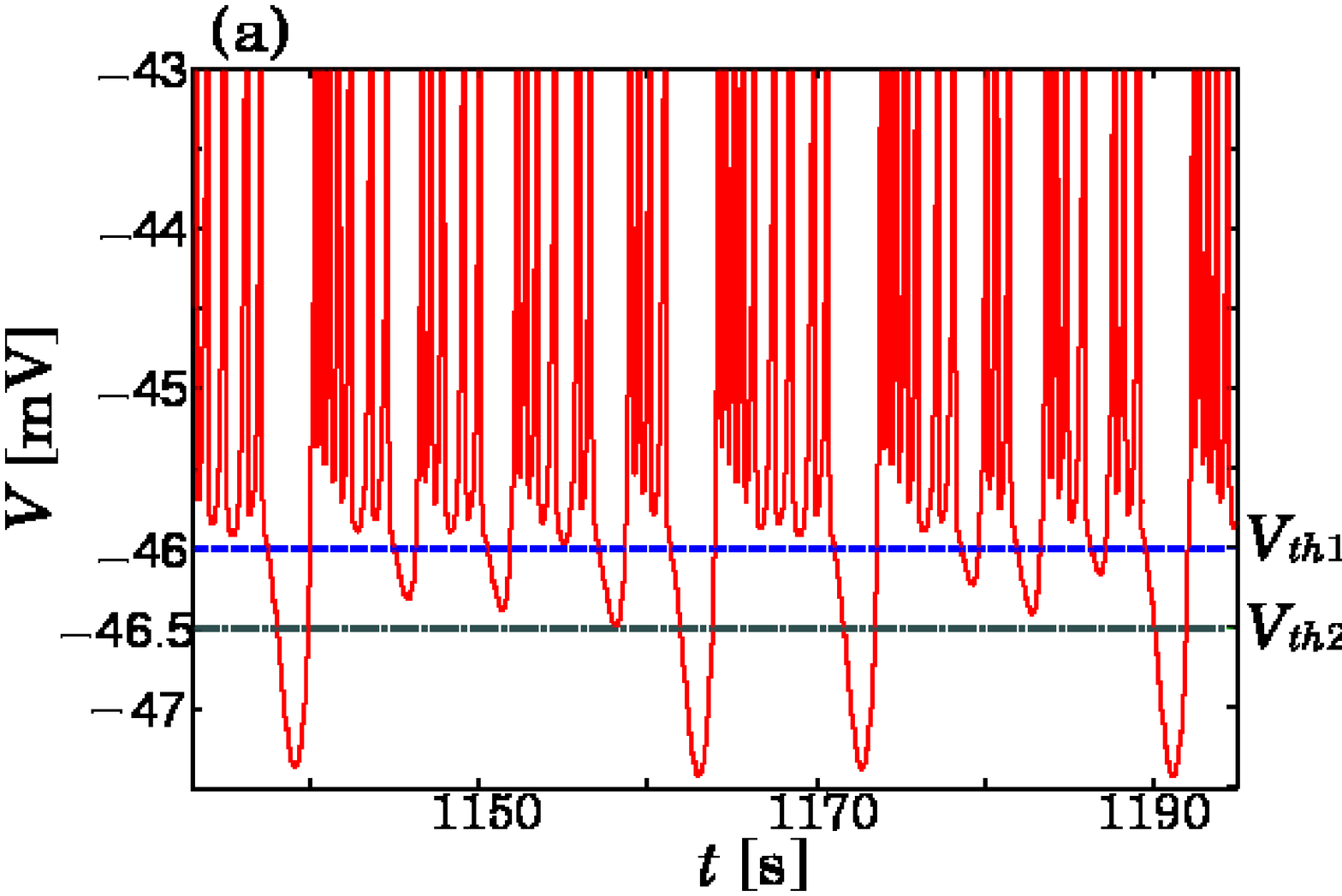}
  \end{center}
  \end{minipage}
  	\begin{minipage}{.45\textwidth}
  \begin{center}
    \includegraphics[keepaspectratio=true,height=45mm]{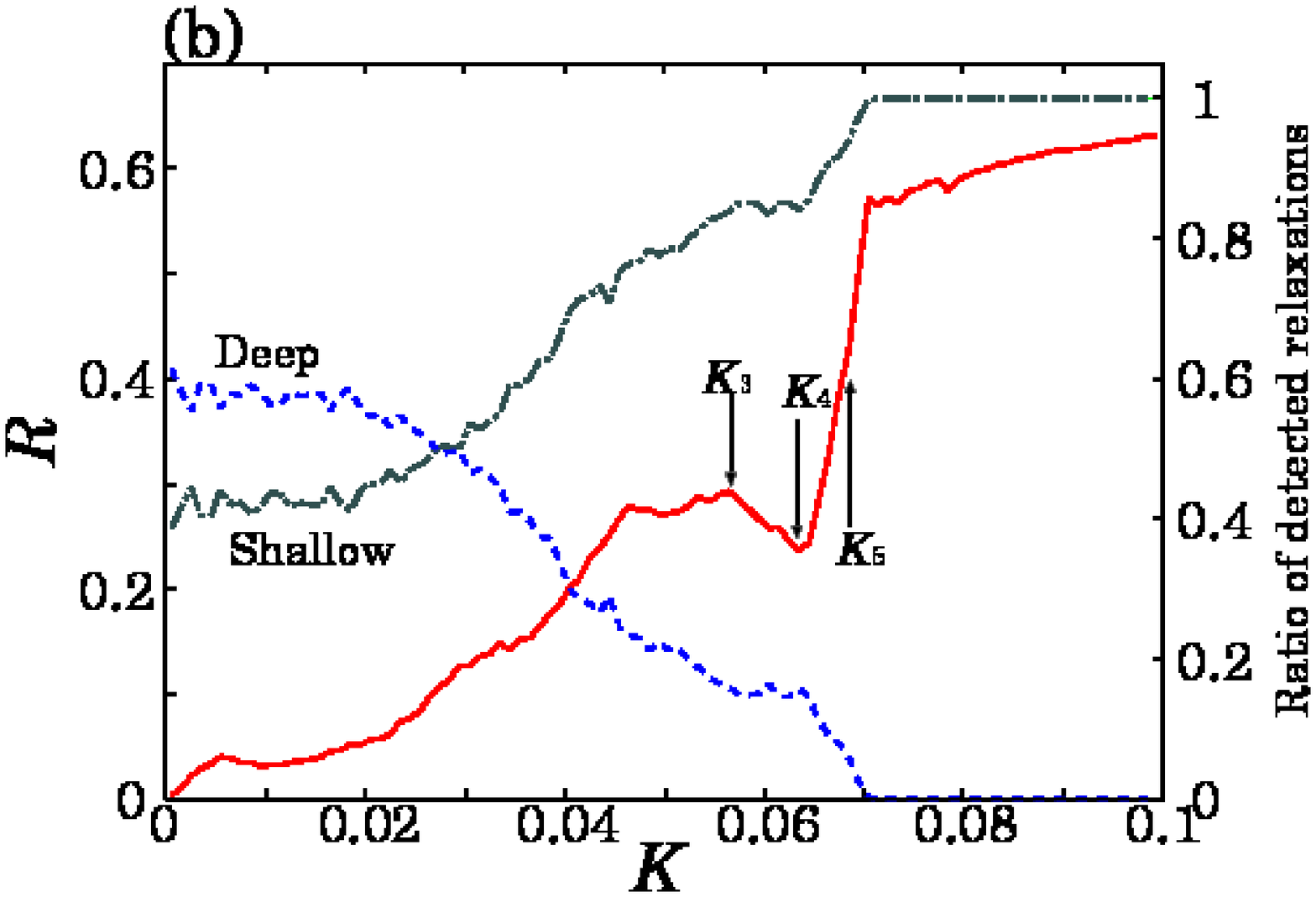}
  \end{center}
  \end{minipage}
  \caption{(Color online) (a) Time series of $V(t)$ 
  with two thresholds, $V_{th1}=-46$ (blue dashed line) and 
  $V_{th2}=-46.5$ (green dashed-dotted line), for $K=0$. 
  (b) {The ratio of shallow (green dashed-dotted line)  and 
  deep  (blue dashed line)  {hyperpolarizations}}  
  with respect to $K$. { The results for (b) are 
  averaged over 100 different realizations.} Also shown is $R$ (red solid line). 
 } 
  \label{fig:Times_relax_o9.eps}
\end{figure}

{{In what follows}, we explain how $R$ increases toward the 
step region in Fig. \ref{fig:worefracOP.eps}(b) and 
how {it} does not {increase} in the step region $\Gamma$.  
These phenomena can be explained {on the basis of} the changes {in}  peaks 
in the probability distribution $P(\theta_n)$ as {follows}.  

First, we investigate {the} {quiescent periods in terms of {the extent in the decrease in} $V$, i.e., {the extent of} hyperpolarization,} with {an} increase {in} $K$.  
As shown in Fig. \ref{fig:Times_relax_o9.eps}(a), 
we {observe} two types of hyperpolarizations in {the} {quiescent periods}, 
namely{,} shallow and deep hyperpolarizations{,}
{ {when} $V{\approx} -46$}.  
{Shallow} hyperpolarizations are {observed}  between $V=-46$ and $-46.5$, 
{whereas} deep {hyperpolarizations} are {observed} below $V=-46.5$. 
Therefore, the two types of hyperpolarizations are 
distinguished by two appropriate thresholds for $V$, {i.e., 
$V_{th1}=-46$ and $V_{th2}=-46.5$}{,} as shown in 
Fig. \ref{fig:Times_relax_o9.eps}(a).  
{The threshold} $V_{th1}$ detects both {the} shallow and {the} deep  
hyperpolarizations, {whereas the} threshold $V_{th2}$ detects only 
deep hyperpolarization. 
Figure \ref{fig:Times_relax_o9.eps}(b) shows {the ratio of 
 shallow and deep hyperpolarizations over all the {hyperpolarizations}  
 detected by $V_{th1}$}  
as a function of $K$, when counted in the interval of 
$1000$ s and summed over $100$ {different realizations}.}

{\subsection*{$R$ on the increase toward the step region $\Gamma$}}

\begin{figure}[htbp]
\begin{minipage}{.45\textwidth}
  \begin{center}
    \includegraphics[keepaspectratio=true,height=50mm]{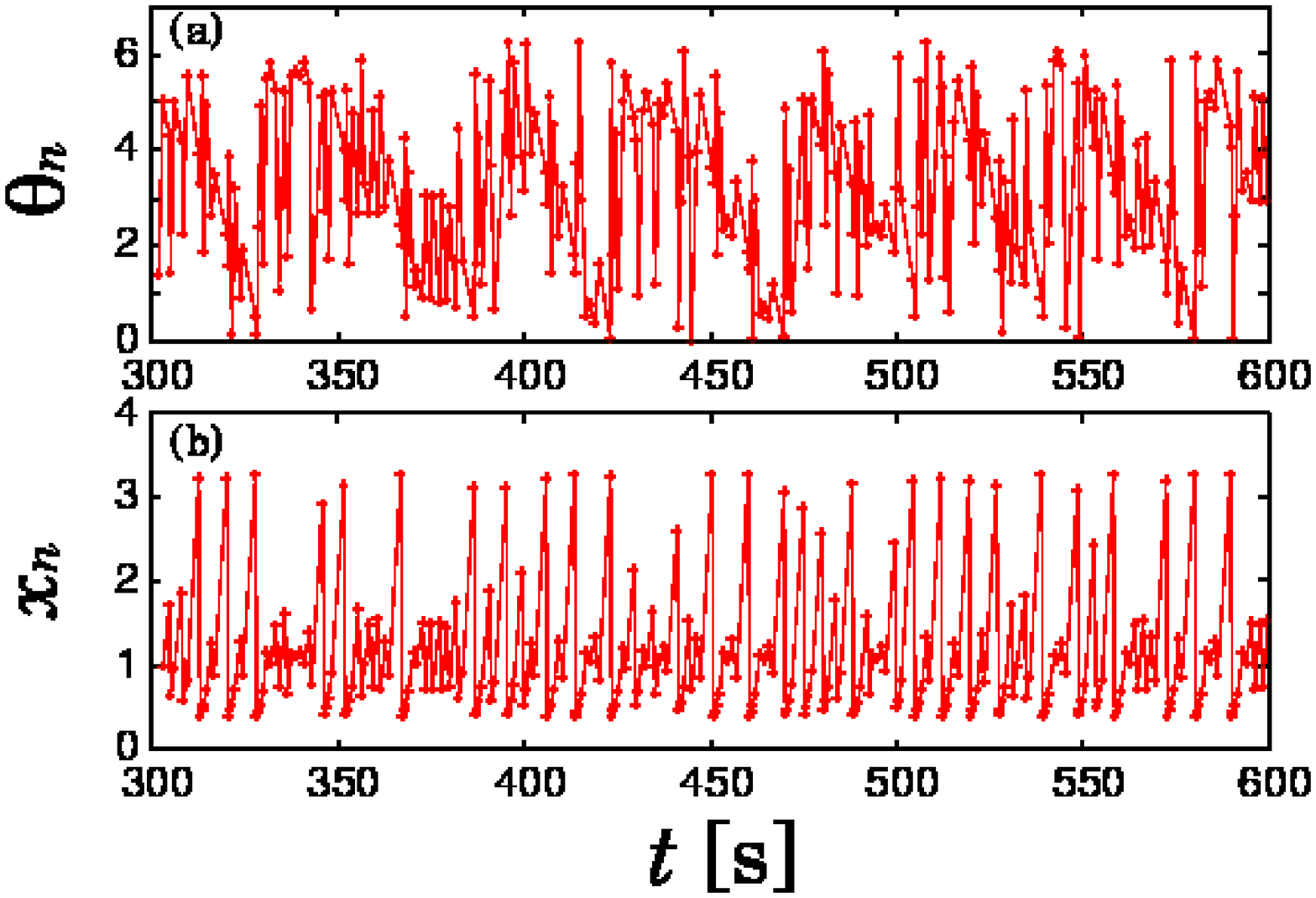}
  \end{center}
  \end{minipage}
  \begin{minipage}{.45\textwidth}
  \begin{center}
    \includegraphics[keepaspectratio=true,height=50mm]{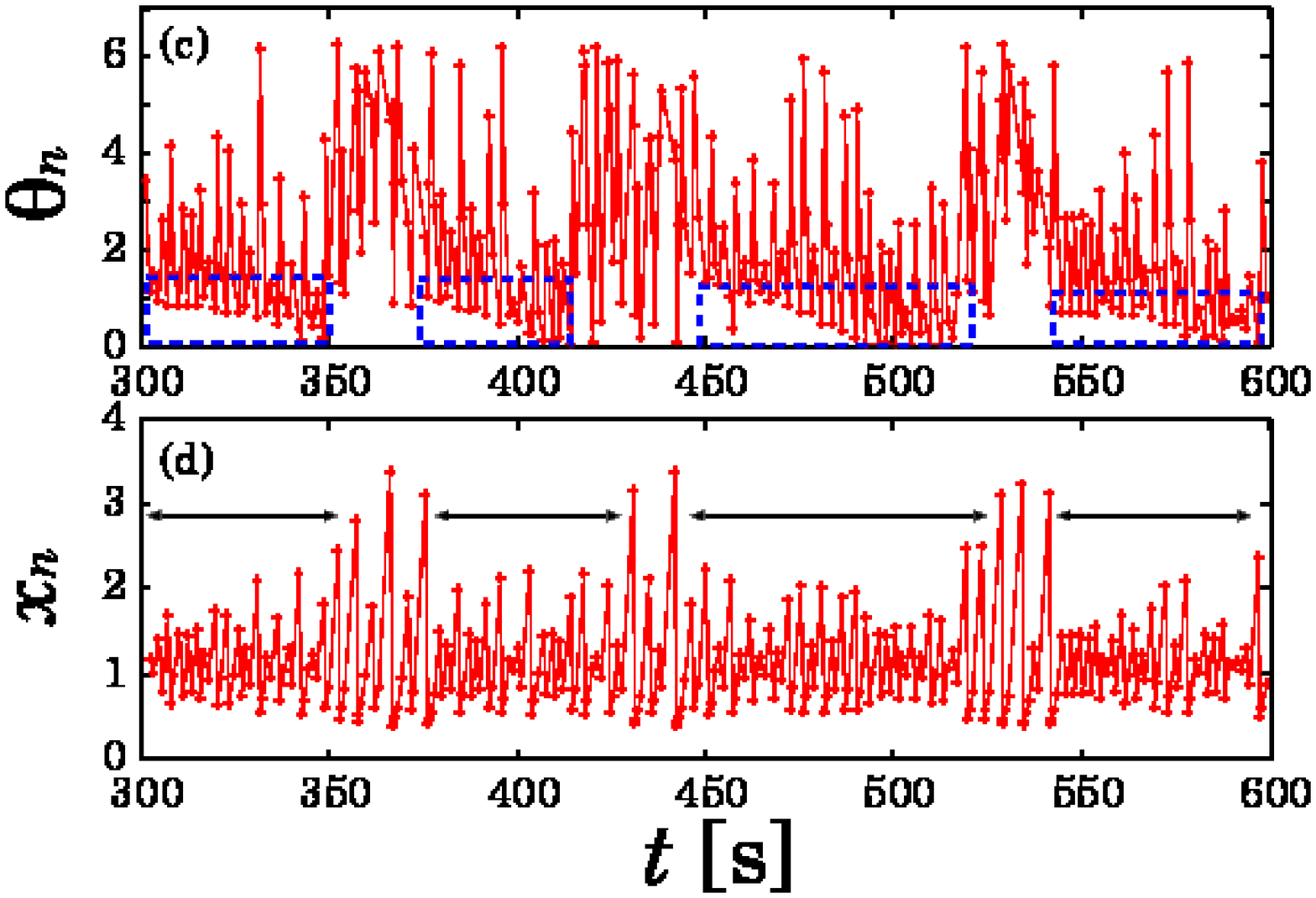}
  \end{center}
  \end{minipage}
  \caption{Time series of $\theta_n$ (a)((c))  and $x_n$ (b)((d)) for $K=0$ ($K=0.05$). The regions indicated by the double-headed arrows in (d) correspond to the bursting periods other than the period of deep hyperpolarizations.}
  \label{fig:ts_thetaISIK005o9.eps}
\end{figure}

{
As shown in Fig. \ref{fig:Times_relax_o9.eps}(b), the number of deep hyperpolarizations  
(blue dashed line) decreases with {an} increase {in} $K$. This {implies} that 
the number of longer ISIs corresponding to deep hyperpolarizations
decreases{,} and this {decrease} is related to 
 an emergence of a peak in $P(\theta_n)${,} as
explained below.   

Let $x_n$ be the length of the ISI between the $n$th 
spike and the $(n+1)$th spike. 
Then, {Figs}. \ref{fig:ts_thetaISIK005o9.eps}(a)((c)) and (b)((d)) show {the} time series of $\theta_n$ and $x_n$ for $K=0$ ($K=0.05$), respectively. 
In Fig. \ref{fig:ts_thetaISIK005o9.eps}(c), we observe gradually shifting 
envelopes for the oscillations of $\theta_n${,} as indicated by the dotted rectangles, 
{whereas} no {such} specific tendency {is observed} in Fig. \ref{fig:ts_thetaISIK005o9.eps}(a). 
The gradual shifts {in} the envelope occur during the time intervals other than  
{those during which} deep {hyperpolarization} {is observed}, which are indicated by the double{-}headed arrows 
in Fig. \ref{fig:ts_thetaISIK005o9.eps}(d). {It should be noted} that deep hyperpolarization corresponds 
to the oscillations of $x_n$ exceeding $x_n\approx 2.5$ and {that} Fig. \ref{fig:ts_thetaISIK005o9.eps}(b) 
{shows a greater number of deep hyperpolarizations} than {those shown} in (d). 

These observations give rise to {a} change in the shape of $P(\theta_n)$ in Fig. \ref{fig:P_theta_o9_2.eps}. {That is}, $P(\theta_n)$ is uniformly distributed for $K=0${,} {owing} to the 
 fluctuation of $\theta_n$ with {a few gradually shifting envelopes}{,} as shown in Fig. \ref{fig:ts_thetaISIK005o9.eps}(a). 
 On the other hand, $P(\theta_n)$ has a peak for $K=0.05${, attributed} to the gradually shifting envelopes{,} 
as shown in Fig. \ref{fig:ts_thetaISIK005o9.eps}(c). The region for the envelopes is around 
$\theta_n \approx 1$ and corresponds to the peak of $P(\theta_n)$ 
in Fig. \ref{fig:P_theta_o9_2.eps}. This appearance of {the} peak in $P(\theta_n)$ 
is {attributed} to the decrease {in} the number of deep hyperpolarizations{,}  
as seen {in} {Figs.} \ref{fig:ts_thetaISIK005o9.eps}(b) and (d). 
}

{\subsection*{$R$ on a plateau in the step region $\Gamma$}} 

\begin{figure}[htbp]
  %\begin{minipage}{.45\textwidth}
  \begin{center}
    \includegraphics[keepaspectratio=true,height=50mm]{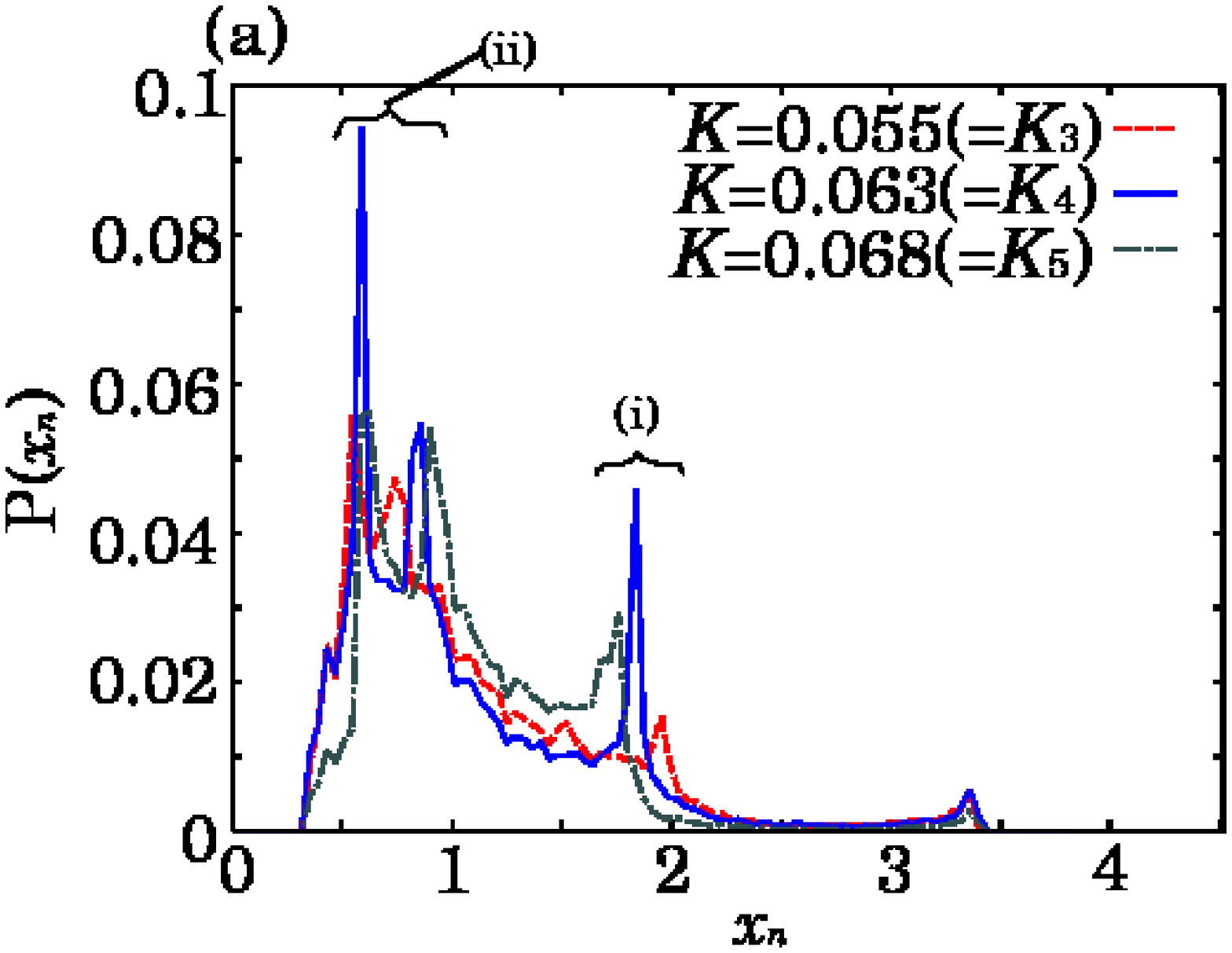}
  \end{center}
  %\end{minipage}
  %\begin{minipage}{.45\textwidth}
  \begin{center}
    \includegraphics[keepaspectratio=true,height=50mm]{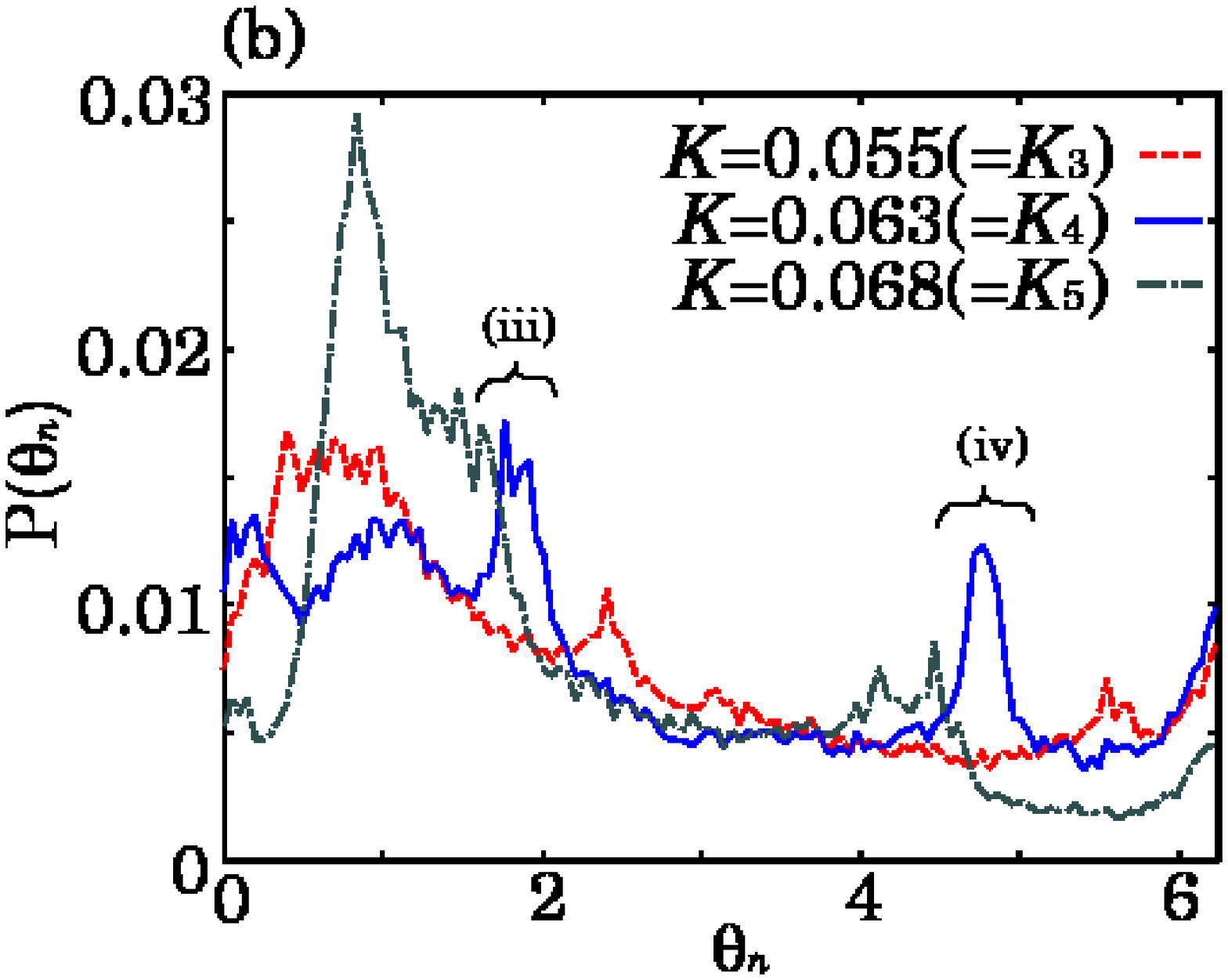}
  \end{center}
  %\end{minipage}
  \caption{(Color online) (a) Probability density distribution of ISIs normalized with respect to their length $x_n$ for $K=0.055\ (=K_3)$ (red dashed line), ${K=}\ 0.063\ (=K_4)$ (blue solid line){,} and ${K=}\ 0.068\ (=K_5)$ (green dashed-dotted line). (b) Probability density distribution of $\theta_n$ corresponding  to (a). }
  \label{fig:P_ISI_o9_2.eps}
\end{figure}

We explain {how} $R$ as a function of 
$K$ remains flat in $\Gamma$. Figure \ref{fig:Times_relax_o9.eps}(b)  
shows that the number of shallow {hyperpolarizations} increases 
with {an} increase {in} $K$ in $\Gamma${,} as well {(green dashed-dotted line)}. 
In order to investigate the increase {in} shallow {hyperpolarizations} in $\Gamma$,  
we consider the distribution of {the} ISIs.  
Figure \ref{fig:P_ISI_o9_2.eps}(a) shows the distribution of ISIs 
for $K=0.055,0.063$, and 
$0.068$, denoted by $K_3,K_4$, and $K_5$, respectively. In {this} figure, 
we can observe peaks located at $x_n{\in[1.5; 2]}$, 
as indicated by (i) in Fig. \ref{fig:P_ISI_o9_2.eps}(a).  
These peaks correspond to shallow {hyperpolarizations} 
and become prominent at the right edge of $\Gamma$. 

\begin{figure}[t]
  \begin{center}
    \includegraphics[keepaspectratio=true,height=45mm]{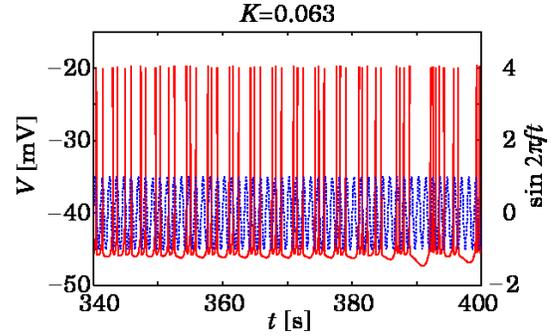}
  \end{center}
  \caption{(Color online) Time series of $V(t)$ (red solid line)  with the 
  periodic forcing (blue dotted line)  for $K=0.063$. }
  \label{fig:ts_K0063_o9.eps}
\end{figure}

Furthermore, for each $K${,} the distribution of ISIs has 
two other peaks at smaller values of $x_n$,  
as indicated by (ii) in Fig. \ref{fig:P_ISI_o9_2.eps}(a).  
These peaks can be  {inferred} from the time series of 
voltage $V(t)$ for a value of $K$ in $\Gamma${,} as shown 
in Fig. \ref{fig:ts_K0063_o9.eps}. As can be seen 
in the interval of $t$ between $350$ { s} and $380$ { s}, 
there exists {a} transient     
$1\ :\ 3$ phase synchronization between one burst 
and three periods of the force.  
The peaks located near $x_n{\in[1.5; 2]}$ in Fig. 
\ref{fig:P_ISI_o9_2.eps}(a) correspond to the inter-burst intervals 
during such a transient phase synchronization. 
{In} contrast, the two other peaks at small $x_n$ 
indicated by (ii) in Fig.  \ref{fig:P_ISI_o9_2.eps}(a) 
correspond to the two ISIs for the three consecutive 
spikes in each burst. 

We also calculate the corresponding distributions  
of $\theta_n${,} as shown in Fig. \ref{fig:P_ISI_o9_2.eps}(b). 
The peaks in the distribution of $x_n$ are related to the peaks 
in $P(\theta_n)$.  
The second and the third spikes 
in each burst  during the transient 
$1\ :\ 3$ phase synchronization correspond to the peaks in $P(\theta_n)$ around 
$\theta_n\  {\approx} \ 2$  
and $\theta_n \ {\approx}\  5$  ((iii) and (iv) in Fig. \ref{fig:P_ISI_o9_2.eps}(b)),  
respectively. 
It should be noted that  
for $K=K_3$, the distribution of $P(\theta_n)$ is concentrated,  
with a large peak (unimodal) located near $\theta_n {\in[0.5; 1]}$. 
Then, for $K=K_4${,} that peak is weakened 
and other peaks (multimodal) appear near $\theta_n \ {\approx} \ 2$ and $\theta_n \ {\approx}\  5$. 

 The shift {in} the peak in the distribution of $P(\theta_n)$ 
 from unimodal to multimodal does not significantly 
 influence the amplitude of the {average} of the STPs 
 whose distribution on the unit circle reflects the shape of 
 $P(\theta_n)$. Therefore, the value of the order parameter 
 $R$ {does not increase} in $\Gamma$.  
However, as $K$ increases further, the peak 
 located at $\theta_n \  {\approx} \ 1$ becomes higher{,}  
 while $R$ increases steeply, and the transition to CPS occurs.

Finally, looking at Figs. \ref{fig:P_ISI_o9_2.eps}(a) and \ref{fig:P_ISI_o9_2.eps}(b)  
in more detail,  {we find that} the peaks in both $P(x_n)$ and $P(\theta_n)$  
for $K_4$ are sharper than  
those for $K_3$ and $K_5$. 
The sharpness of the peaks corresponds to the duration of 
the transient $1\ :\ 3$ phase synchronization. 
In fact, the contribution to $R$ from the newly emerged  
peaks in $P(\theta_n)$ around $\theta_n \ {\approx}\  2$ and $\theta_n \ {\approx} \ 5$ 
 is the {greatest}  
when $K\ {\approx}\  K_4$. {On the other hand}, the contribution of 
the first concentrated peak around $\theta \ {\approx}\  0.5$ {decreases} at that value of $K$. 
As a consequence, even though $K$ increases, 
 { which would normally cause the} degree of synchronization {to} increase, 
the value of $R$ decreases near $K=K_4$. 
A similar phenomenon has been reported as anomalous 
phase synchronization, whereby coupling among interacting 
oscillator systems {increases} the natural frequency disorder 
before synchronization \cite{BlasiusPRE03}.  

%%%Appendix till here%%%%

%\bibliography{cc}

\begin{thebibliography}{99}
\bibitem{footnote}By intensive simulations for $\langle \tau_{sl}\rangle$ in the region 
where 
 parameter $K$ is closer to the transition point, we are able to
{numerically observe}
another power-law scaling that is similar to { 
the case of the transition to chaotic phase synchronization 
for a system {that} has a single rotation center{,} with only 
one characteristic time scale.}   
However, further investigations are needed to confirm whether this    
scaling corresponds to the conventional case.
\end{thebibliography}

\end{document}